\documentclass[aps,showpacs,preprintnumbers,amsmath, amssymb,nofootinbib]{revtex4}

\usepackage{float}
\usepackage{graphics,epsfig}
\usepackage{bm}
\usepackage{slashed}
\usepackage{graphicx}
\usepackage{amsmath}
\usepackage{amsfonts}
\usepackage{amssymb}
\usepackage{color}%
\usepackage{dcolumn}
\providecommand{\U}[1]{\protect\rule{.1in}{.1in}}
\begin{document}
\baselineskip=0.8 cm \title{Dipole Coupling Effect of Holographic Fermion in the Background of Charged Gauss-Bonnet AdS Black Hole }
\author{Xiao-Mei Kuang$^{1}$}
\email{xmeikuang@gmail.com}
\author{Bin Wang$^{1}$}
\email{wang_b@sjtu.edu.cn}
\author{Jian-Pin Wu$^{2}$}
\email{jianpinwu@mail.bnu.edu.cn}
\affiliation{$^{1}$ INPAC, Department of Physics and Shanghai
Key Lab for Particle Physics and Cosmology,
Shanghai Jiao Tong University, Shanghai 200240,
China\\
$^{2}$ Department of Physics, Beijing Normal University, Beijing 100875, China }

\vspace*{0.2cm}
\begin{abstract}
\baselineskip=0.6 cm
\begin{center}
{\bf Abstract}
\end{center}
We investigate the  holographic fermions in the charged Gauss-Bonnet $AdS_{d}$ black hole background with the dipole coupling between fermion and gauge field in the bulk. We show that in addition to the strength of the dipole coupling, the spacetime dimension and the higher curvature correction in the gravity background also influence the onset of the Fermi gap and the gap distance. We find that the higher curvature effect modifies the fermion spectral density and influences the  value of the Fermi momentum for the appearance of the Fermi surface. There are richer physics in the boundary fermion system due to the modification in the bulk gravity.
\end{abstract}

\pacs{11.25.Tq, 04.50.Gh, 71.10.-w}\maketitle
\newpage
\vspace*{0.2cm}

\section{Introduction}

The AdS/CFT correspondence\cite{Maldacena,Gubser-1,E.Witten} is a great achievement in string theory. It has opened new avenues for studying the strongly-coupled many body phenomena by relating certain interacting quantum field theories to classical gravity systems.
Recently stimulated by this correspondence, a remarkable connection between the condensed matter and the gravitational physics has been discovered, for reviews see \cite{Hartnoll,herzog,horowitz}.
It was first suggested in \cite{Gubser-2,Gubser-3} that near the horizon of a charged black hole there is in operation a geometrical mechanism parameterized by a charged scalar field of breaking a local $U(1)$ gauge symmetry. This spontaneous $U (1)$ symmetry breaking by bulk black holes can be used to construct gravitational duals of the transition from normal state to superconducting state in the boundary theory. The gravity models with the property of holographic superconductor have attracted considerable interest for their potential applications to the condensed matter physics.

It is of interest to consider a quantum field theory which contains fermions charged under a global $U(1)$ symmetry. Fermionic sectors possess a number of generic features which might lead to interesting phenomena related to condensed matter physics. However, many of them have not been discussed in the available holographic studies. When a finite $U(1)$ charged density in the fermionic sector is introduced in the holographic system, it is natural to ask whether the system possesses a Fermi surface and what is the low energy excitations. There have been some progresses in studying the fermionic sector, where a number of generic couplings for the fermions have been realized\cite{Sung-Sik Lee,HongLiuUniversality,HongLiuNon-Fermi,HongLiuSpinor,FiniteT,HongLiuAdS2,BTZ,Magnetic1,Magnetic2,
Magnetic3,Leigh1a,Leith2a,HFlatBand,FlatBandDipole,LifshitzFermions1,Fang}. Recently, introducing the coupling between the fermion and gauge field through a dipole interaction in the bulk, it was remarkably found that as the strength of the interaction is varied, spectral density is transferred and beyond a critical interaction strength a gap opens up\cite{R.G.Leigh1}. The existence of Fermi surfaces as the varying of the dipole coupling was also disclosed\cite{R.G.Leigh2}. The extended investigation on the dipole coupling also can be seen in \cite{JPWu2,Wen, JPWu3}.

Most studies on the gravitational constructions of the holographic superconductors are based on the Einstein gravity background. It would be interesting to see how the modification of the bulk gravity background may influence the property in the condensation on the boundary. Considering that the string theory contains higher curvature corrections in the gravity arising from stringy effects, it is intriguing to examine the higher curvature correction effect on the holographic superconductor. From the AdS/CFT correspondence, the higher curvature corrections on the gravity side will lead corrections in the boundary field theory. In studying the spontaneous $U (1)$ symmetry breaking with charged scalar field coupling to the gauge field, it was found that the higher curvature correction can make the condensation harder to form and influence the properties in conductivity and other properties of condensations\cite{GBHS1,GBHS2,GBHS3,GBHS4,GBHS5,GBHS6,GBHS7}. The  higher curvature influence in the holographic fermion system when fermion is minimally coupled to the gauge field was also examined in \cite{JPWu1}. The main motivation of the present paper is to further study the effect of the higher curvature correction in the bulk gravity on the holographic fermion system when there is dipole coupling between the fermion and gauge fields. We are going to investigate how the spacetime dimension and the higher curvature correction in gravity modifies the properties of Fermi gap, Fermi momentum etc. in the Fermi system when there is dipole interaction between fermion and gauge fields.

The organization of this paper is as follows. In section II, we set up the formalism describing the equation of motion in the fermionic system in the bulk d-dimensional Gauss-Bonnet charged AdS black hole. In section III, we investigate the influences on the Fermi gaps, Fermi surfaces due to the spacetime dimension and the Gauss-Bonnet factor when there is dipole coupling between fermion and gauge fields.  Finally in section IV, we give the conclusions and discussions.

\section {equations of motion in the bulk}\label{CBHinGBG}

We consider the non-minimal coupling between a spin-1/2 fermions and the gauge field in the form of the dipole interaction described by the bulk action
\begin{eqnarray}
\label{actionspinor}
S_{D}=i\int d^{d}x \sqrt{-g}\overline{\zeta}\left(\Gamma^{a}\mathcal{D}_{a} - m - ip \slashed{F} \right)\zeta,
\end{eqnarray}
where $m$ is the mass of the fermion field, $p$ is the strength of the dipole coupling. In the action,
$\Gamma^{a}=(e_{\mu})^{a}\Gamma^{\mu}$, $\slashed{F}=\frac{1}{4}\Gamma^{\mu\nu}(e_\mu)^a(e_\nu)^bF_{ab}$ and $\mathcal{D}_{a}=\partial_{a}+\frac{1}{4}(\omega_{\mu\nu})_{a}\Gamma^{\mu\nu}-iqA_{a}$,
with  $\Gamma^{\mu\nu}=\frac{1}{2}[\Gamma^\mu,\Gamma^\nu]$ and  the spin connection $(\omega_{\mu\nu})_{a}=(e_\mu)^b\nabla_a(e_\nu)_b$, where
$(e_\mu)^{a}$ forms a set of orthogonal normal vector bases \cite{Conventions1}.

We will concentrate on the  $d$ dimensional charged black hole in Gauss-Bonnet gravity for the bulk configuration, which has the metric \cite{GBblackhole1,GBblackhole2}
\begin{eqnarray}\label{MetricA}
ds^{2}&=&-g_{tt}dt^2+g_{rr}dr^2+g_{xx} \sum_{i=1}^{d-2}(dx^i)^2 \nonumber\\
&=&-f(r)dt^{2}+\frac{dr^{2}}{f(r)}+\frac{r^{2}}{L_{\rm eff}^2} \sum_{i=1}^{d-2}(dx^i)^2
\end{eqnarray}
with
\begin{eqnarray}\label{Leff}
L^2_{\rm eff}=\frac{2\alpha}{1-\sqrt{1-\frac{4\alpha}{L^2}}}
\to  \left\{
\begin{array}{rl}
L^2   \ , &  \quad {\rm for} \ \alpha \rightarrow 0 \\
\frac{L^2}{2}  \ , &  \quad {\rm for} \  \alpha \rightarrow \frac{L^2}{4}
\end{array}\right.
\,
\end{eqnarray}
describing the effective radius of the AdS space in the Gauss-Bonnet gravity.  The gauge connection is written as $A_{a}=A_{t}(r)(dt)_{a}$, where
\begin{eqnarray}
\label{metricA}
A_{t}=\mu\left(1-\frac{r_{+}^{d-3}}{r^{d-3}}\right).
\end{eqnarray}
The metric coefficient reads\footnote{We set the gravitational constant $\kappa_{d}^{2}=1/2$
, the AdS radius $L=1$ and the effective dimensionless gauge field coupling parameter $g_{F}=2$.}
\begin{eqnarray}
\label{metricf}
f(r)=\frac{r^{2}}{2\alpha}\left[1-\sqrt{1-4\alpha\left(1-\frac{r_{+}^{d-1}}{r^{d-1}}\right)+\frac{2(d-3)\alpha \mu^{2}r_{+}^{d-3}}{(d-2)r^{d-1}}\left(1-\frac{r_{+}^{d-3}}{r^{d-3}}\right)}\right].
\end{eqnarray}
$r_{+}$ is the event horizon radius which is characterized by $f(r_{+})=0$ and $\mu$ can be identified with the chemical potential of the dual field theory. The causality gives strong constraint on the Gauss-Bonnet coupling \cite{GBcouplingConstraint1,GBcouplingConstraint2,GBcouplingConstraint3,GBcouplingConstraint4,GBcouplingConstraint5,
GBcouplingConstraint6,GBcouplingConstraint7,GBcouplingConstraint8,GBcouplingConstraint9}
to be in the range $-\frac{7}{36}\leq \alpha\leq \frac{9}{100}$.
It is easy to check that in the limit $\alpha\rightarrow 0$, (\ref{metricf}) goes back to the form for the Reissner-Nordstr$\ddot{o}$m
AdS black hole.
The Hawking temperature of the charged Gauss-Bonnet AdS black hole reads
\begin{eqnarray}
\label{HawkingT}
T=\frac{f'(r_{+})}{4\pi}=\frac{(d-1)r_{+}}{4 \pi}\left(1-\frac{(d-3)^2 \mu^{2}}{2(d-1)(d-2)r_{+}^2}\right),
\end{eqnarray}
which can be viewed as the temperature of the conformal field theory on the AdS boundary. Note that we will set $r_+=1$ in the following investigation.

Now we can write down the Dirac equation of the fermions in the bulk spacetime.
To go to the momentum space,  we transform
$\zeta=(-g g^{rr})^{-\frac{1}{4}}F e^{-i\omega t +ik_{i}x^{i}}$ and set $k_i=k \delta_{i}^{1}$ without loss of generality. Then the Dirac  equation has the form
\begin{eqnarray}
\label{DiracEinFourier}
(\sqrt{g^{rr}}\Gamma^{r}\partial_{r}- m
- \frac{i p}{2} \sqrt{g^{rr}g^{tt}} \Gamma^{rt} \partial_{r}A_{t})F
-i(\omega+q A_{t})\sqrt{g^{tt}}\Gamma^{t}F
+i k \sqrt{g^{xx}}\Gamma^{x}F=0,
\end{eqnarray}
It is obvious that (\ref{DiracEinFourier}) only depends on three Gamma matrices $\Gamma^{r},\Gamma^{t},\Gamma^{x}$.
So it is convenient to express $F$ into $F=(F_{1},F_{2})^{T}$ and
choose the following basis for our gamma matrices \cite{Photoemission}:
\begin{eqnarray}
\label{GammaMatrices}
 && \Gamma^{r} = \left( \begin{array}{cc}
-\sigma^3 \textbf{1}  & 0  \\
0 & -\sigma^3 \textbf{1}
\end{array} \right), \;\;
 \Gamma^{t} = \left( \begin{array}{cc}
 i \sigma^1 \textbf{1}  & 0  \\
0 & i \sigma^1 \textbf{1}
\end{array} \right),  \;\;
\Gamma^{x} = \left( \begin{array}{cc}
-\sigma^2 \textbf{1}  & 0  \\
0 & \sigma^2 \textbf{1}
\end{array} \right),
\qquad \ldots
\end{eqnarray}
The Dirac equation can be rewritten into
\begin{eqnarray} \label{DiracEF}
\sqrt{g^{rr}}\partial_{r}\left( \begin{matrix} F_{1} \cr  F_{2} \end{matrix}\right)
+m\sigma^3\otimes\left( \begin{matrix} F_{1} \cr  F_{2} \end{matrix}\right)
=\sqrt{g^{tt}}(\omega+qA_{t})i\sigma^2\otimes\left( \begin{matrix} F_{1} \cr  F_{2} \end{matrix}\right)
\mp  k \sqrt{g^{xx}}\sigma^1 \otimes \left( \begin{matrix} F_{1} \cr  F_{2} \end{matrix}\right) \nonumber \\
-p \sqrt{g^{tt}g^{rr}}\partial_{r}A_{t}\sigma^1\
\otimes \left( \begin{matrix} F_{1} \cr  F_{2} \end{matrix}\right).
\end{eqnarray}

Furthermore, we
will set $F_{I} = ( \mathcal{A}_{I} ,  \mathcal{B}_{I})^{T}(I=1,2)$  to decouple the equation of motion.
Under such decomposition, the Dirac equation (\ref{DiracEF}) can be divided into
\begin{eqnarray} \label{DiracEAB1}
(\sqrt{g^{rr}}\partial_{r}\pm m)\left( \begin{matrix} \mathcal{A}_{1} \cr  \mathcal{B}_{1} \end{matrix}\right)
=\pm(\omega+qA_{t})\sqrt{g^{tt}}\left( \begin{matrix} \mathcal{B}_{1} \cr  \mathcal{A}_{1} \end{matrix}\right)
-(k \sqrt{g^{xx}}+p \sqrt{g^{tt}g^{rr}}\partial_{r}A_{t})\left( \begin{matrix} \mathcal{B}_{1} \cr  \mathcal{A}_{1} \end{matrix}\right)
~,
\end{eqnarray}
\begin{eqnarray} \label{DiracEAB2}
(\sqrt{g^{rr}}\partial_{r}\pm m)\left( \begin{matrix} \mathcal{A}_{2} \cr  \mathcal{B}_{2} \end{matrix}\right)
=\pm(\omega+qA_{t})\sqrt{g^{tt}}\left( \begin{matrix} \mathcal{B}_{2} \cr  \mathcal{A}_{2} \end{matrix}\right)
+(k \sqrt{g^{xx}}-p \sqrt{g^{tt}g^{rr}}\partial_{r}A_{t}) \left( \begin{matrix} \mathcal{B}_{2} \cr  \mathcal{A}_{2} \end{matrix}\right)
~.
\end{eqnarray}

It is convenient to introduce $\xi_{I}\equiv \frac{\mathcal{A}_{I}}{\mathcal{B}_{I}}(I=1,2)$ and reduce the Dirac equations (\ref{DiracEAB1}) and (\ref{DiracEAB2})
into the non-linear flow equation
\begin{eqnarray} \label{DiracEF1}
(\sqrt{f(r)}\partial_{r}+2m)\xi_{I}=\left[ v_{-} + (-1)^{I} k \frac{L_{\rm eff}}{r} \right]
+ \left[ v_{+} - (-1)^{I} k \frac{L_{\rm eff}}{r}  \right]\xi_{I}^{2}~
\end{eqnarray}
where $v_{\pm}=\frac{1}{\sqrt{f(r)}}\left[\omega+q\mu\left(1-\frac{1}{r^{d-3}}\right)\right]\pm(d-3) p \mu\frac{1}{r^{d-2}}$.

We will numerically solve the Dirac equation by imposing the boundary condition. Near the AdS boundary, from (\ref{DiracEF}) we see that the reduced Dirac field behaves as
\begin{eqnarray} \label{BoundaryBehaviour}
F_{I} \buildrel{r \to \infty}\over {\approx} a_{I}r^{-m L_{\rm eff}}\left( \begin{matrix} 1 \cr  0 \end{matrix}\right)
+b_{I}r^{m L_{\rm eff}}\left( \begin{matrix} 0 \cr  1 \end{matrix}\right),
\qquad
I = 1,2~.
\end{eqnarray}
As discussed in \cite{HongLiuSpinor,HongLiuAdS2}, if $a_{I}\left( \begin{matrix} 1 \cr  0 \end{matrix}\right)$
and $b_{I}\left( \begin{matrix} 0 \cr  1 \end{matrix}\right)$ are related by
$a_{I}\left( \begin{matrix} 1 \cr  0 \end{matrix}\right)
=\mathcal{S}b_{I}\left( \begin{matrix} 0 \cr  1 \end{matrix}\right)$,
then the boundary Green's functions $G(\omega,k)$ is given by $G=-i \mathcal{S}\gamma^{0}$.
The Green's functions can be expressed in the form
\begin{eqnarray} \label{GreenFBoundary}
G (\omega,k)=
\left( \begin{array}{cc}
G_{11}(\omega,k)\textbf{1}   & 0  \\
0  & G_{22}(\omega,k)\textbf{1} \end{array} \right)  \ =\lim_{r\rightarrow \infty} r^{2m L_{\rm eff}}
\left( \begin{array}{cc}
\xi_{1}\textbf{1}   & 0  \\
0  & \xi_{2}\textbf{1} \end{array} \right)  \ .
\end{eqnarray}
Solving the flow equation (\ref{DiracEF1}) with the boundary condition at the horizon
\begin{eqnarray} \label{GatTip}
\xi_{I}\buildrel{r \to 1}\over =i,
\end{eqnarray}
we can get the Green function $G_{II}(\omega,k)$.

When the background becomes extremal, the metric coefficient behaves as $f(r)\sim(d-1)(d-2)(r-1)^2$ near the horizon. This makes taking the limit $\omega\rightarrow 0$ near the horizon subtle, in which the geometry approaches $AdS_{2}\times \mathbb{R}^{d-2}$
\begin{equation}\label{ads2}
ds^2=\frac{1}{(d-1)(d-2)\varsigma^2}(-d\tau^2+d\varsigma^2)+\frac{1}{L_{\rm eff}^2}\sum_{i=1}^{d-2}(dx^i)^2
\end{equation}
for $T=0$ with $\varsigma=\frac{\omega L^2}{(d-1)(d-2)(r-r_+)}$ and $\tau=\omega t$. In this region we can
expand the Dirac field $F$ in terms of $\varsigma$ in powers of $\omega$ as
\begin{equation}\label{match1}
\left( \begin{matrix} F_{1}(\varsigma) \cr  F_{2}(\varsigma) \end{matrix}\right)=
\left( \begin{matrix} F_{1}^{(0)}(\varsigma) \cr  F_{2}^{(0)}(\varsigma) \end{matrix}\right)+\omega \left( \begin{matrix} F_{1}^{(1)}(\varsigma) \cr  F_{2}^{(1)}(\varsigma) \end{matrix}\right)+\omega^2 \left( \begin{matrix} F_{1}^{(2)}(\varsigma) \cr  F_{2}^{(2)}(\varsigma) \end{matrix}\right)+\cdots.
\end{equation}
By substituting (\ref{match1}) into (\ref{DiracEF}), we have the leading order term
\begin{eqnarray}\label{matchequation}
\partial_{\varsigma}\left( \begin{matrix} F_{1}^{(0)}(\varsigma) \cr  F_{2}^{(0)}(\varsigma) \end{matrix}\right)=\frac{1}{\sqrt{(d-1)(d-2)}\varsigma}m \sigma^{3}
\left( \begin{matrix} F_{1}^{(0)}(\varsigma) \cr  F_{2}^{(0)}(\varsigma) \end{matrix}\right)-i(1+\frac{(d-3)q\mu}{(d-1)(d-2)\varsigma})\sigma^{2}\left( \begin{matrix} F_{1}^{(0)}(\varsigma) \cr  F_{2}^{(0)}(\varsigma) \end{matrix}\right)\\ \nonumber
+\frac{1}{\sqrt{(d-1)(d-2)}\varsigma}[(d-3)p\mu-(-1)^{I} k L_{\rm eff}]\sigma^{1}\left( \begin{matrix} F_{1}^{(0)}(\varsigma) \cr  F_{2}^{(0)}(\varsigma) \end{matrix}\right)
.
\end{eqnarray}
It is the equation of motion for spinor fields with masses\cite{HongLiuAdS2}
\begin{equation}\label{mtilde}
[m,\tilde{m}_{I}=(d-3)p\mu-(-1)^{I}kL_{\rm eff}]
\end{equation}
in $AdS_2$ background, where $\tilde{m}_{I}(I=1,2)$ are time-reversal violating mass terms. According to the analysis in \cite{HongLiuAdS2}, $F_{I}^{(0)}(\varsigma)$ is dual to the spinor operators $\mathbb{O}_{I}$ in the IR $CFT_{1}$ with the
conformal dimensions $\delta_{I}=\nu_{I}(k)+\frac{1}{2}$ where
\begin{equation}\label{nuk}
\nu_{I}(k)=\sqrt{\frac{m^2+\tilde{m}_{I}^2}{(d-1)(d-2)}-\big[\frac{(d-3)q\mu}{(d-1)(d-2)}\big]^2}
=\frac{\sqrt{2}q}{\sqrt{(d-1)(d-2)}}\sqrt{\frac{m^2+\tilde{m}_{I}^2}{2q^2}-1}~~~~~~~(I=1,2).
\end{equation}
To obtain the second equality, we have used $\mu=\frac{\sqrt{2(d-1)(d-2)}}{d-3}$ for zero temperature. It is obvious that the two coupling parameters $p$ and $\alpha$ imprint the scaling in the IR. By matching the inner $AdS_2$ and outer $AdS_4$ solutions in the matching region where we consider
$\varsigma\rightarrow0$ and $\omega/\varsigma\rightarrow0$\cite{HongLiuAdS2}, we can express the coefficients $a_{I}$ and $b_{I}$ in (\ref{BoundaryBehaviour}) as
\begin{eqnarray}\label{coefficients}
a_{I}&=&[a_{I}^{(0)}+\omega a_{I}^{(1)}+\cdots]+[\tilde{a}_{I}^{(0)}+\omega \tilde{a}_{I}^{(1)}+\cdots] \mathcal{G}_{I}(k,\omega),\nonumber \\
b_{I}&=&[b_{I}^{(0)}+\omega b_{I}^{(1)}+\cdots]+[\tilde{b}_{I}^{(0)}+\omega \tilde{b}_{I}^{(1)}+\cdots] \mathcal{G}_{I}(k,\omega),
\end{eqnarray}
where $a_{I}^{(n)}, \tilde{a}_{I}^{(n)}, b_{I}^{(n)}$ and $\tilde{b}_{I}^{(n)}$ can be determined numerically and $\mathcal{G}_{\alpha}(k,\omega)$
is the retarded Green functions of the dual operators $\mathbb{O}_{I}$ with the form\cite{HongLiuAdS2}
\begin{equation}\label{greenads2}
 \mathcal{G}_{I}(k,\omega)=\left\{e^{-i \pi \nu_{I}(k)}\frac{\Gamma(-2\nu_{I}(k))\Gamma(1+\nu_{I}(k)-i\frac{(d-3)q\mu}{(d-1)(d-2)})
 [\frac{(m+i\tilde{m}_{I})}{\sqrt{(d-1)(d-2)}}-i\frac{(d-3)q\mu}{(d-1)(d-2)}-\nu_{I}(k)]}
{\Gamma(2\nu_{I}(k))\Gamma(1-\nu_{I}(k)-i\frac{(d-3)q\mu}{(d-1)(d-2)})[\frac{(m+i\tilde{m}_{I})}{\sqrt{(d-1)(d-2)}}-i\frac{(d-3)q\mu}{(d-1)(d-2)}+\nu_{I}(k)]}\right\}\omega^{2  \nu_{I}(k)}
\end{equation}
We see that the Gauss-Bonnet coupling and dipole coupling modify the dual Green function via $\tilde{m}_{I}$. It is noticed that (\ref{coefficients}) is only valid when $2\nu_{I}(k)$ is not an integer. In the case when it is an integer, terms like $\omega^n log(\omega)$ should be added \cite{HongLiuAdS2}.

Instead of (\ref{GatTip}), the boundary condition of $\xi_{I}$ for $\omega=0$ is found in the form
\begin{equation}\label{bdyw0}
\xi_{I}\buildrel{r \to 1}\over =\frac{\frac{m}{\sqrt{(d-1)(d-2)}}-\nu_{I}(k)}{\frac{(d-3)q\mu}{(d-1)(d-2)}+\frac{\tilde{m}_{I}}{\sqrt{(d-1)(d-2)}}}.
\end{equation}
Thus, when $\omega=0$,  one should employ the boundary condition  (\ref{bdyw0}) instead of (\ref{GatTip}) to  numerically solve the flow equation (\ref{DiracEF1}).

\section{influences on the fermion system due to the dipole coupling, the spacetime dimension and the Gauss-Bonnet factor}

We numerically integrate the flow equation (\ref{DiracEF1}) and read off the asymptotic values to compute the matrix of the retarded Green functions.  We will calculate the fermion spectral function $A(\omega,k)\equiv \rm{Tr}[\space Im G(\omega,k)]$ and also the density of states $A(\omega)$ by doing the integration of $A(\omega,k)$ over $k$. Furthermore we will investigate the dipole coupling effect in the limit of $\omega=0$ and the existence of the Fermi surfaces.

\subsection{Dipole coupling effect in different dimensional Einstein background}
In this subsection, we will explore the dipole coupling effect in different dimensional background spacetimes.
We will neglect the curvature correction in the bulk by setting $\alpha=0$ for the moment.

\begin{figure}
\center{
\includegraphics[scale=0.3]{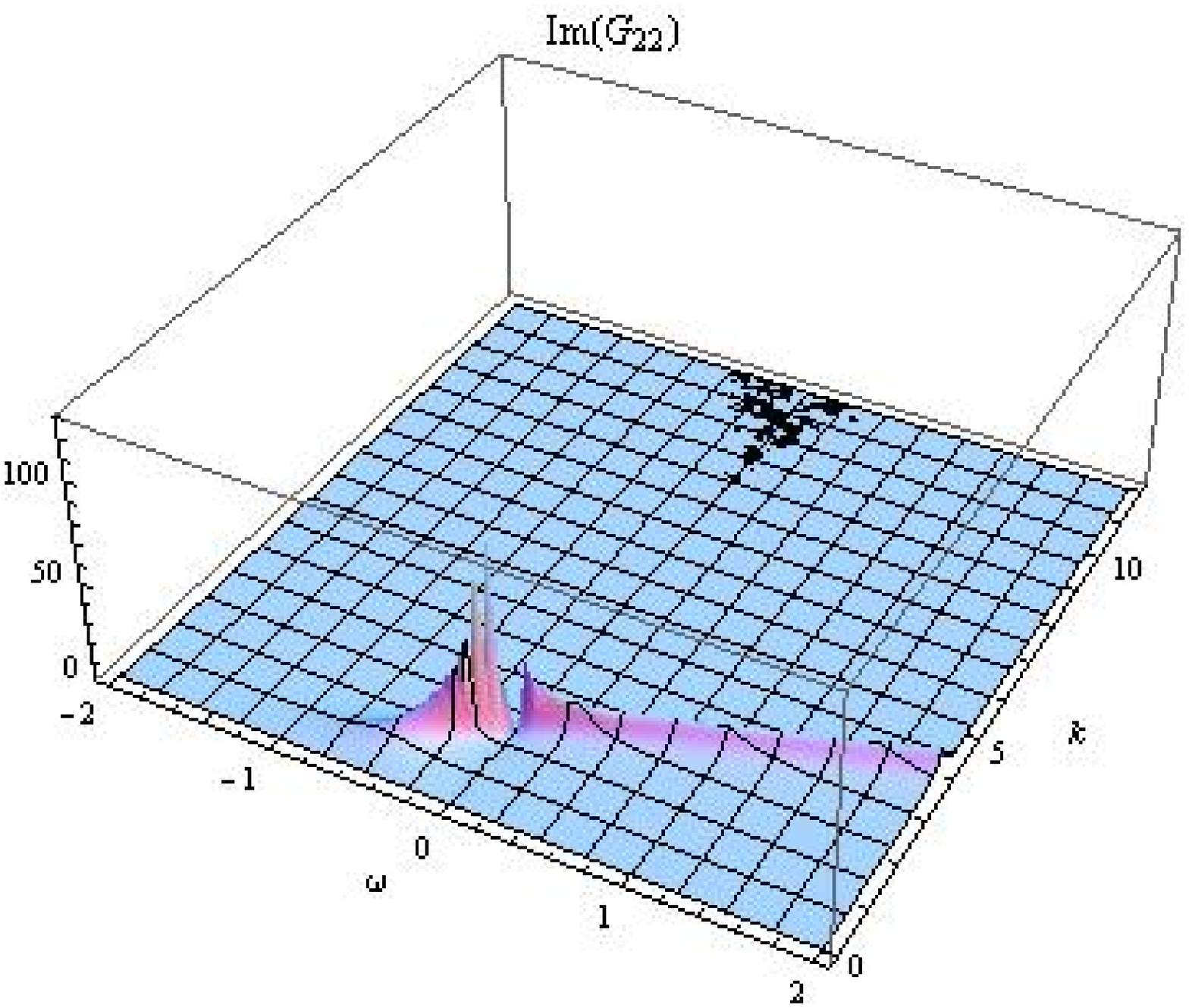}\hspace{1.5cm}
\includegraphics[scale=0.3]{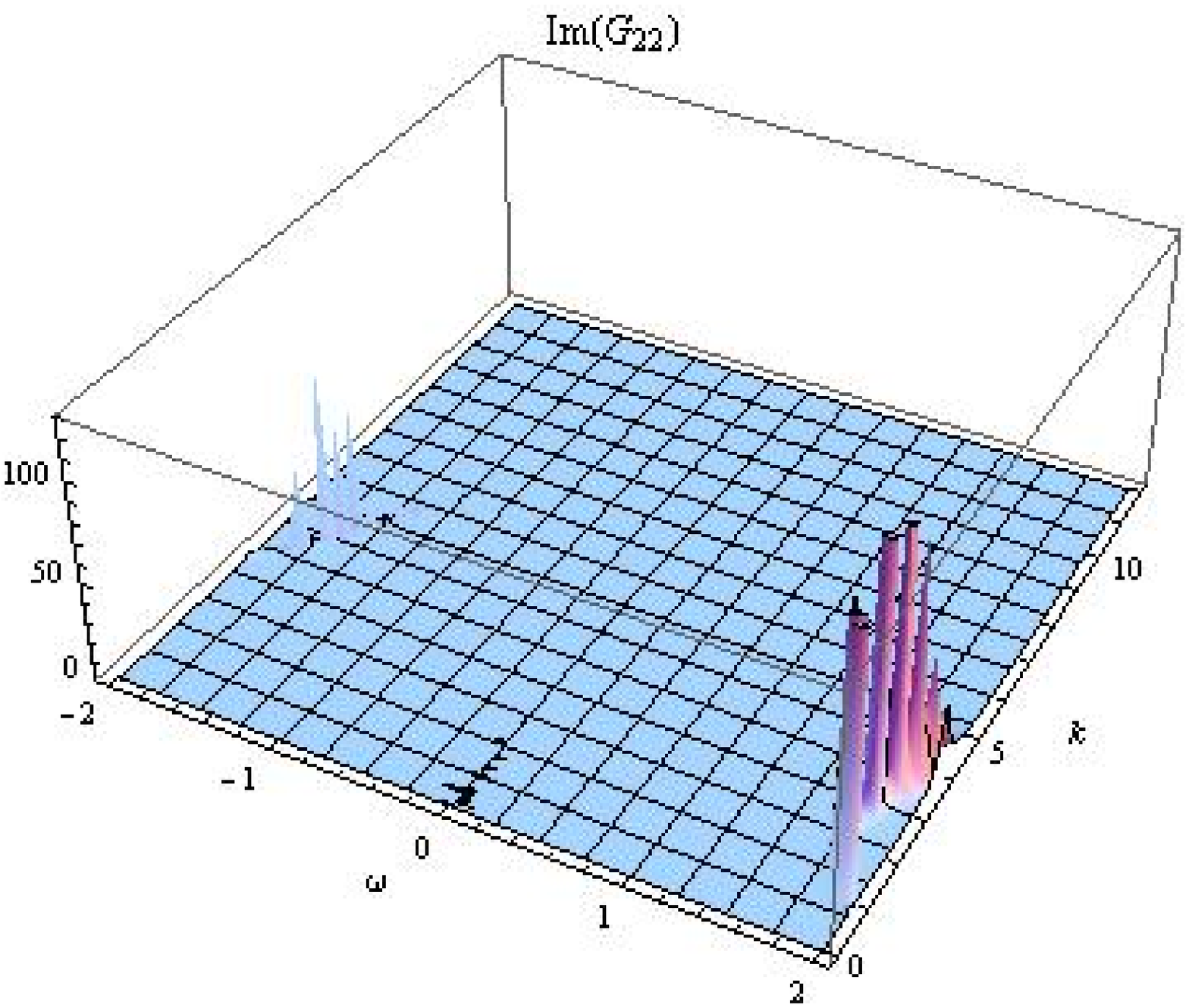}\\
\includegraphics[scale=0.25]{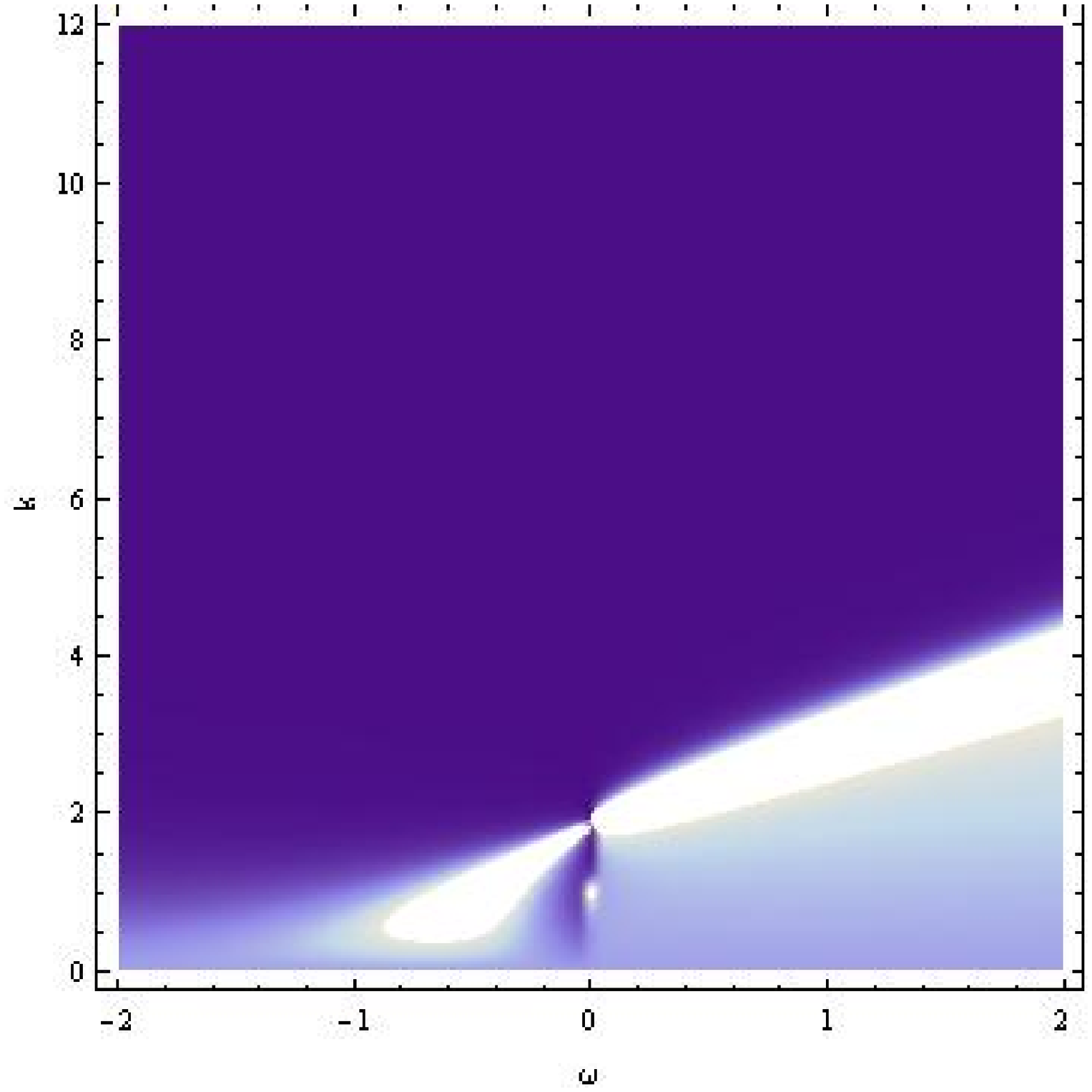}\hspace{1.5cm}
\includegraphics[scale=0.25]{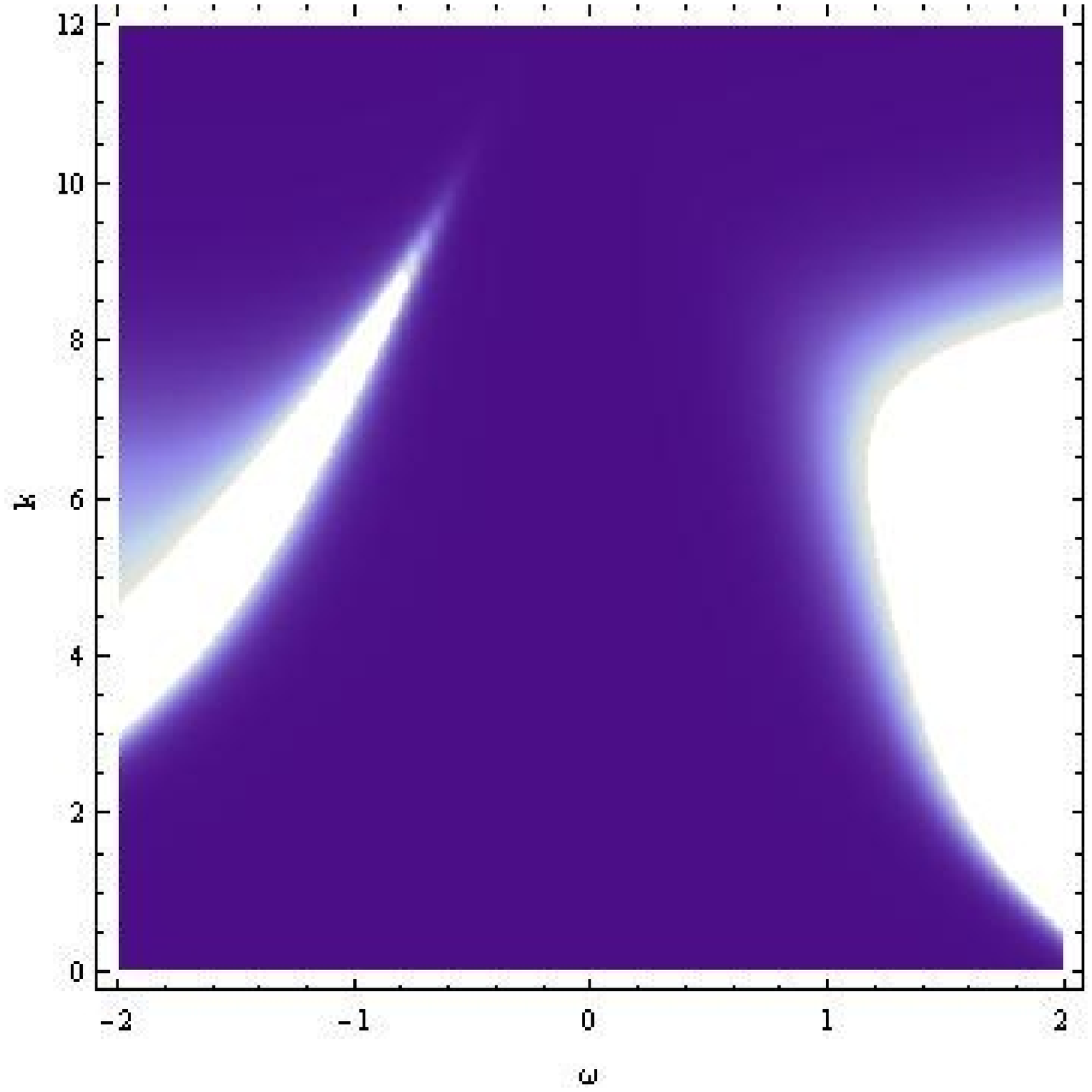}\\
\caption{\label{fig3D1} $\rm{Im} [G_{22}(\omega,k)]$ for $p=0$ (left plane) and $p=3$ (right plane) with $d=5$.}}
\end{figure}

\begin{figure}
\center{
\includegraphics[scale=0.7]{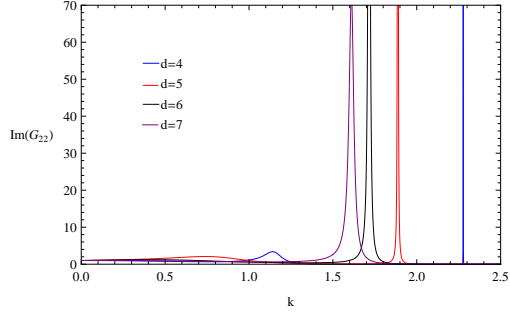}
\caption{\label{kF-d} The plot of $\rm{Im}[G_{22}(\omega,k)]$ for different $d$ dimensional AdS black hole at $p=0$. Here we set $\omega =-10^{-8}$. }}
\end{figure}

First for the minimal dipole coupling with $p=0$, we can discuss the Fermi momentum $k_{F}$, the dispersion relation and disclose the effect of spacetime dimension in the Fermi system. In the left plot of Fig. \ref{fig3D1}, we have reproduced the 3D plot of $\rm{Im[G_{22}(\omega,k)]}$ for $p=0$ disclosed in \cite{JPWu1}. The sharp quasi-particle-like peak at $\omega=0$ represents a Fermi surface. Furthermore, in Fig. \ref{kF-d}, it shows $\rm{Im}[G_{22}(\omega,k)]$ for different spacetime dimension $d$, where we find $k_F$ for the sharp quasi-particle-like peak gets smaller in higher dimensional spacetime. This property holds as well when the dipole coupling is non-minimal.
Improving the accuracy, we determine the Fermi momentums as $2.2769$\footnote{The fermion momentum is different from the value $0.92$ in \cite{HongLiuNon-Fermi} because $g_F$ is set differently. With the same value of parameter as in \cite{HongLiuNon-Fermi} we can reproduce $0.92$.}, $1.8873, 1.7160$ and $1.6106$ for $d=4,5, 6$ and $7$ respectively.
Once $k_F$ is determined, as discovered in \cite{HongLiuSpinor}, the spectral function $\rm{Im}[G_{22}(\omega,k)]$ has a dispersion relation
\begin{eqnarray} \label{LdispersionA}
\tilde{\omega}(\tilde{k})\propto \tilde{k}^{z}, \quad {\rm with} \quad z = \begin{cases} \frac{1}{2 \nu_{I}(k_F)} & \nu_{I}(k_F) < \frac{1}{2}\cr
            1 & \nu_{I}(k_F) > \frac{1}{2}
            \end{cases}.
\end{eqnarray}
where $\tilde{k}=k-k_F$ and $\tilde{\omega}(\tilde{k})$ is the
location of the maximum of the quasi-particle-like peak. Note that $\nu_{I}(k_F)$ has the form in (\ref{nuk}) for $k=k_F$. When $\alpha=0$ and $p=0$, we have $\tilde{m}_{I}=-(-1)^{I}k$. Therefore,
\begin{equation}\label{nu12}
\nu_1(k)=\nu_2(k)=\nu(k)=\frac{\sqrt{2}q}{\sqrt{(d-1)(d-2)}}\sqrt{\frac{m^2+k^2}{2q^2}-1}
\end{equation}
in our model. After determining the Fermi momentum from numerical calculation, we can analytically compute the scaling exponent $z$ of the dispersion relation through (\ref{LdispersionA}) and (\ref{nu12}).
The results are summarized in Table \ref{dkF}. We see that the scaling exponent $z$ of the dispersion relation decreases with the decrease of the spacetime dimension.
\begin{table}
\centering
\begin{tabular}{|c|c|c|c|c|}
  \hline
  $d$ & 4 & 5 & 6 & 7  \\ \hline
  $k_F$ & 2.2769 & 1.8873& 1.7160 & 1.6106 \\ \hline
$z$ & 1 & 1.38591& 2.30064 &3.55325 \\ \hline
\end{tabular}
\caption{The Fermi momentum and scaling exponent of the dispersion relation with different dimensions of background at $p=0$.}
\label{dkF}
\end{table}

Considering the AdS/CFT dictionary where the conformal dimension of the dual fermion operator is $\Delta=\frac{d-1}{2}\pm m L$ in the $d$-dimensional AdS spacetime, we can easily accept the dimensional influence disclosed above. The dimensional effect on the scaling exponent $z$
of the dispersion relation lies in two factors.
The obvious one is the exponent $\nu_{I}(k)$, which depends on the dimension $d$ as shown in (25).
The other is the Fermi momentum $k_{F}$, which is determined by the UV physics and we need to work it out numerically. In general, for RN-AdS background, it depends on the charge $q$ and dimension $\Delta$
(for $m=0$, equivalently the spacetime dimension $d$). Although we can not give a general analytical expression for $k_{F}$, there is an allowed range  for $k_{F}$ \cite{HongLiuAdS2}
\begin{equation}\label{kFrange}
\frac{d-3}{\sqrt{(d-1)(d-2)}}\leq\frac{k_{F}}{\mu_{q}}\leq 1,
\end{equation}
where the lower limit is obviously related to the spacetime dimension $d$. Thus the dimensional influence is quite intrinsic.

\begin{figure}
\center{
\includegraphics[scale=0.7]{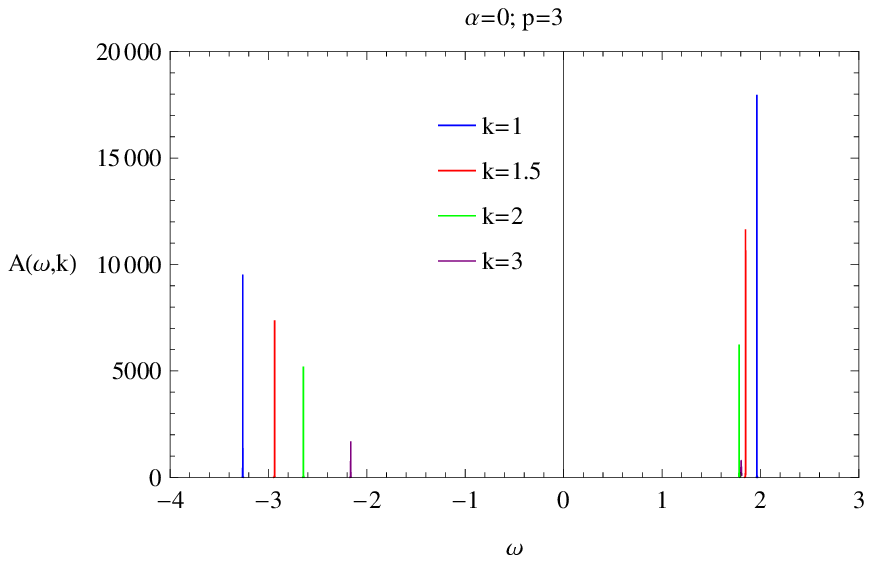}
\includegraphics[scale=0.7]{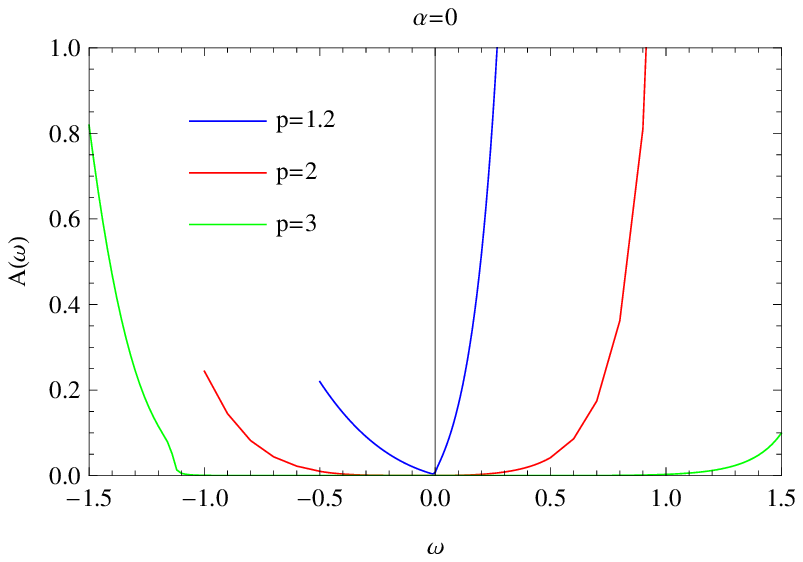}
\caption{\label{figAalpah0} Left: $A(\omega, k)$ as a function of $\omega$ for some values of k at $p=3$. Right: $A(\omega)$ as function of $\omega$ for $p = 1.2$, $2$ and $3$. The onset of the gap for $\alpha=0$ is at $p=2$.}}
\end{figure}

\begin{figure}
\center{
\includegraphics[scale=0.7]{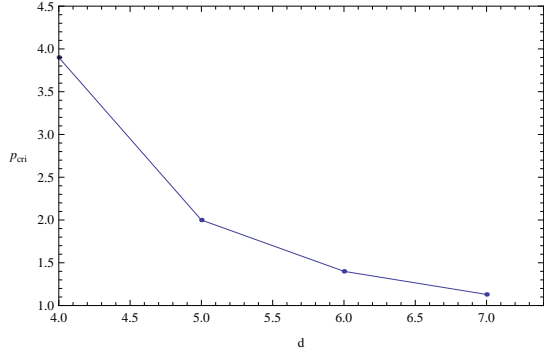}\\
\caption{\label{figdpcri} The critical $p_{cri}$ for the onset of gap versus $d$ for $\alpha=0$.}}
\end{figure}

Now, we turn on the dipole coupling.
Look at the right plots in Fig. \ref{fig3D1}, for $p=3$, instead of a sharp quasi-particle-like peak at $\omega=0$,
we see that a gap opens near $\omega=0$.
The gap in the spectral density exists for all $k$ as shown in the left plot of Fig. \ref{figAalpah0}.
To further determine the onset of the gap, we calculate $A(\omega)$ which is shown in the right part of Fig. \ref{figAalpah0}
\footnote{$A(\omega)$ is the total spectral weight. In numerical calculations, similar to \cite{R.G.Leigh1,R.G.Leigh2,JPWu2}, what we do is to compute $A(\omega,k)$ for various $\omega$ over a sufficiently wide range of $k$, and took the appropriate area under that curve. Then we repeat it for other values of $\omega$. We defined the gap to correspond to the point where the spectral weight drops below some small number, which is approximately $10^{-9}-10^{-8}$ in this paper. For comparison with the results in \cite{R.G.Leigh1}, we have repeated some numerical results, where the small number is also approximately $10^{-9}-10^{-8}$.}. We find that the gap opens at $p=2$ in our 5-dimensional background. Our critical value of $p$ for the onset of gap is different from that discussed in \cite{R.G.Leigh1} for the 4-dimensional background. In Fig. \ref{figdpcri}, we show the influence on the critical $p$ by spacetime dimensions.
It is clear that with the increase of the spacetime dimension, the smaller dipole coupling can make the gap appear. The analytical map between the spacetime dimension $d$ and the critical value of $p$ is lacking, however from the expression of $v_{\pm}$ in the flow equation (\ref{DiracEF1}) we see that $d$ and $p$ are closely related by the product $(d-3)p$. The dipole interaction strength $p$ makes the gap open and plays the role of
the interaction strength in terms of the Hubbard model\cite{R.G.Leigh1}. The spacetime dimension $d$ can influence the product $(d-3)p$ in the flow equation so that can compensate the effect of $p$. This explains why for higher dimension, even smaller critical $p$ can make the gap open.

\begin{figure}
\center{
\includegraphics[scale=0.7]{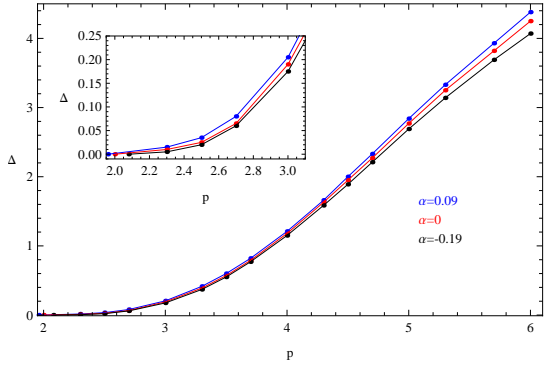}\\
\caption{\label{figgap} The width of gap as a function of $p$ and $\alpha$ for $d=5$.}}
\end{figure}

In Fig. \ref{figgap}, we plot the width of the gap versus $p$ for the chosen spacetime dimension. It is obvious that when we neglect the curvature correction in the bulk spacetime, further increase of the dipole coupling $p$ can lead the gap to become wider. This property keeps in different dimensional configuration.


\subsection{Dipole coupling effect in  Gauss-Bonnet gravity}

\begin{figure}
\center{
\includegraphics[scale=0.3]{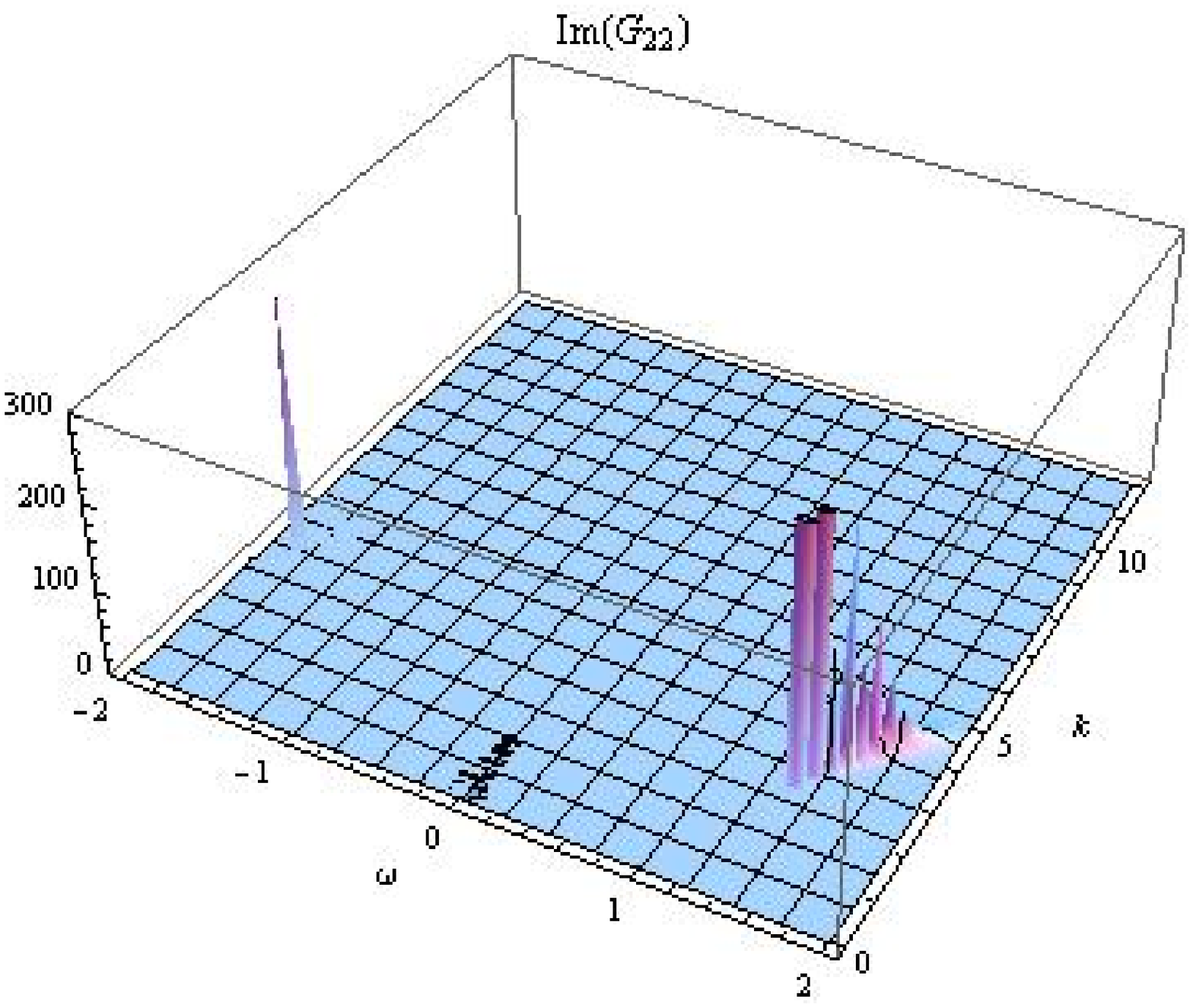}\hspace{1.5cm}
\includegraphics[scale=0.25]{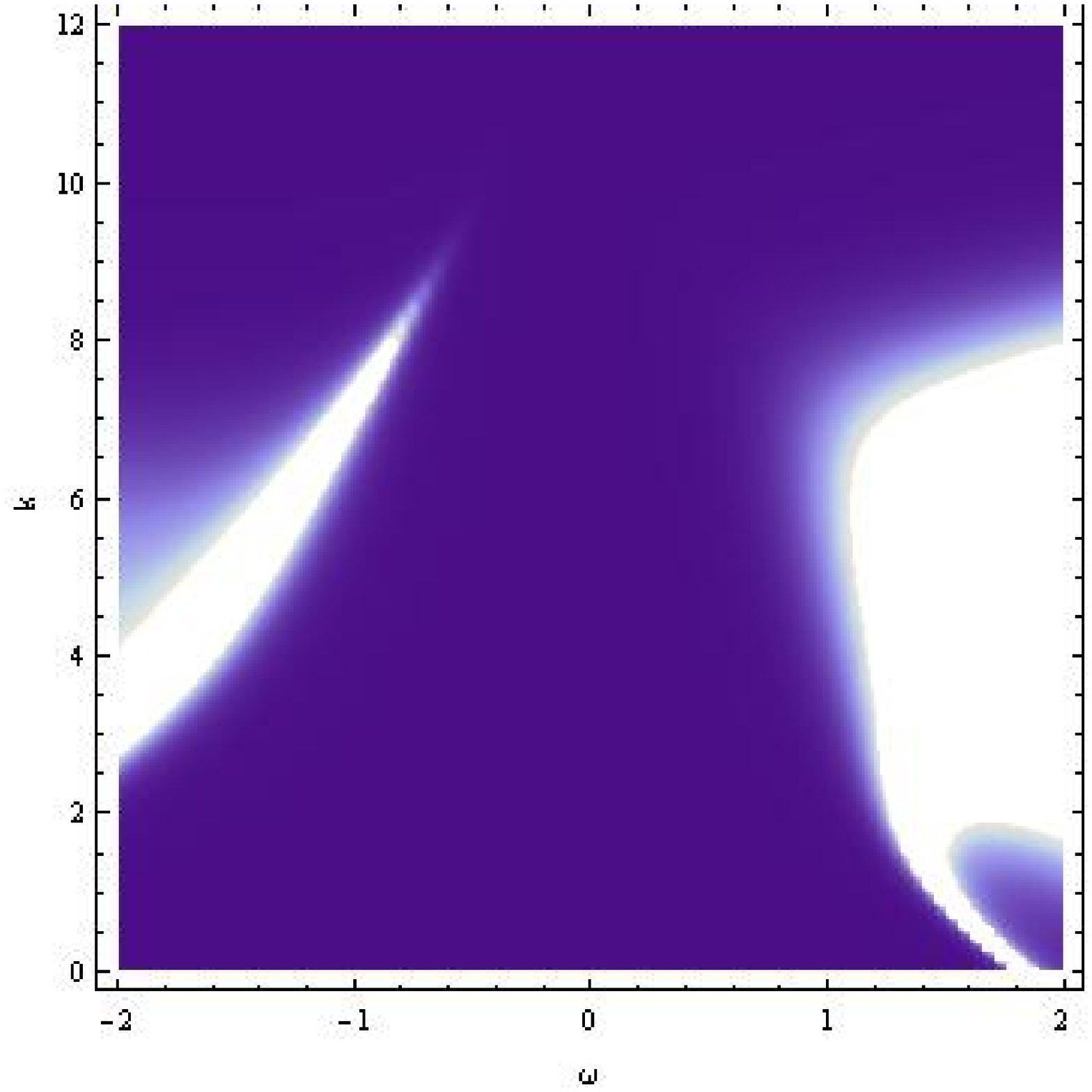}\\
\includegraphics[scale=0.3]{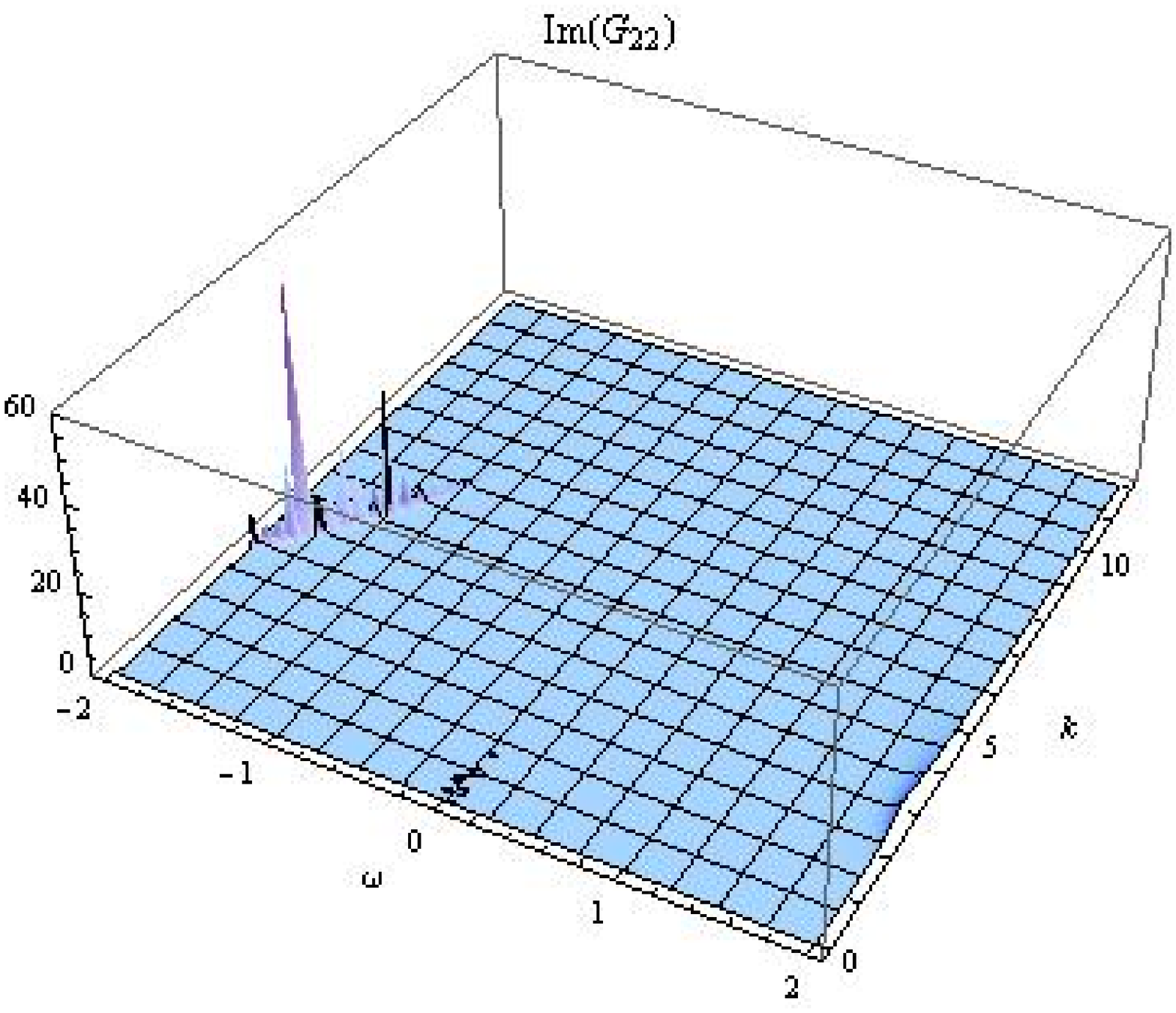}\hspace{1.5cm}
\includegraphics[scale=0.25]{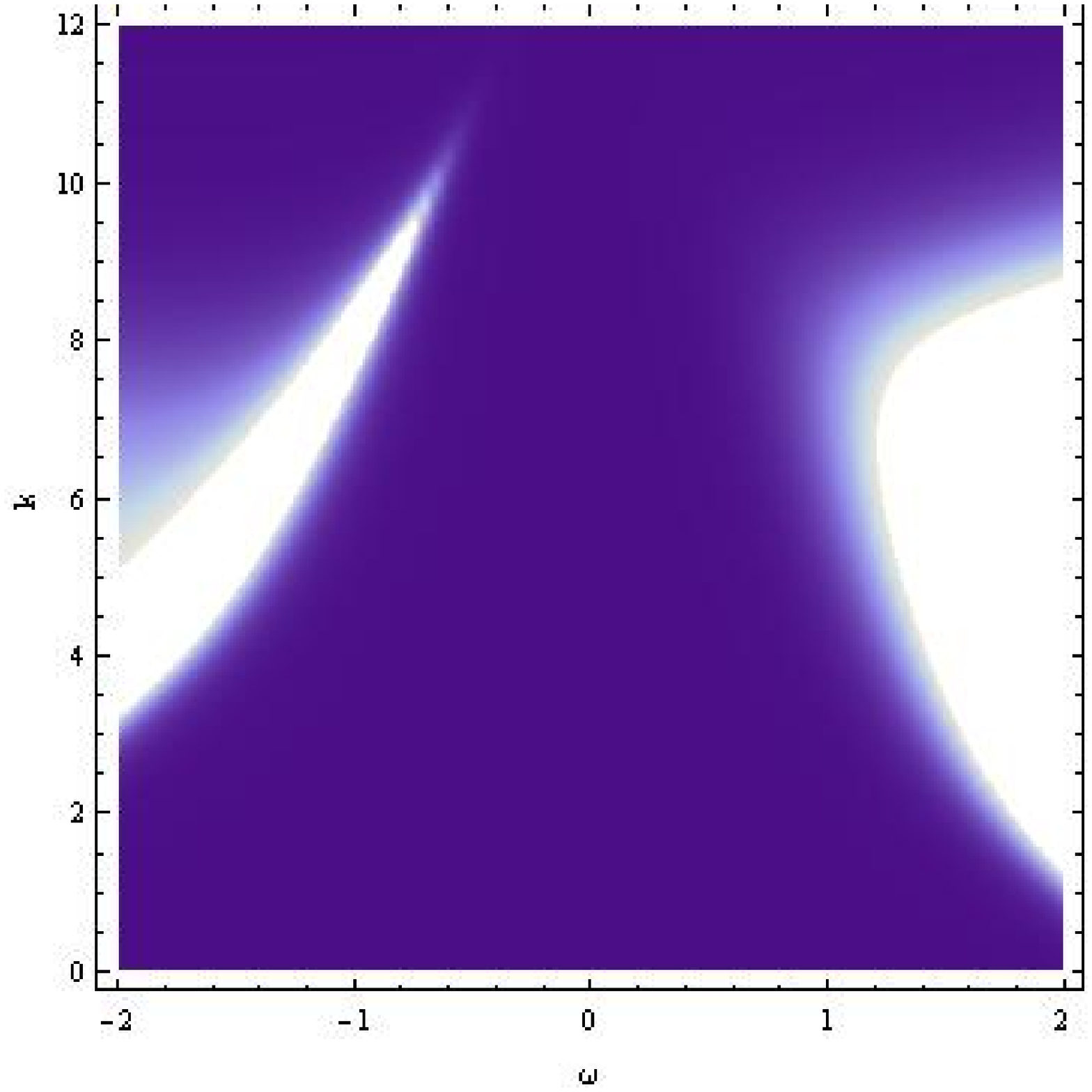}\\
\caption{\label{fig3D2} $\rm{Im} G_{22} (\omega,k)$ for $p=3$. The plots from up to bottom are for $\alpha=-0.19$ and $0.09$. For $p=3$, the gap exists for both chosen $\alpha$.}}
\end{figure}

\begin{figure}
\center{
\includegraphics[scale=0.7]{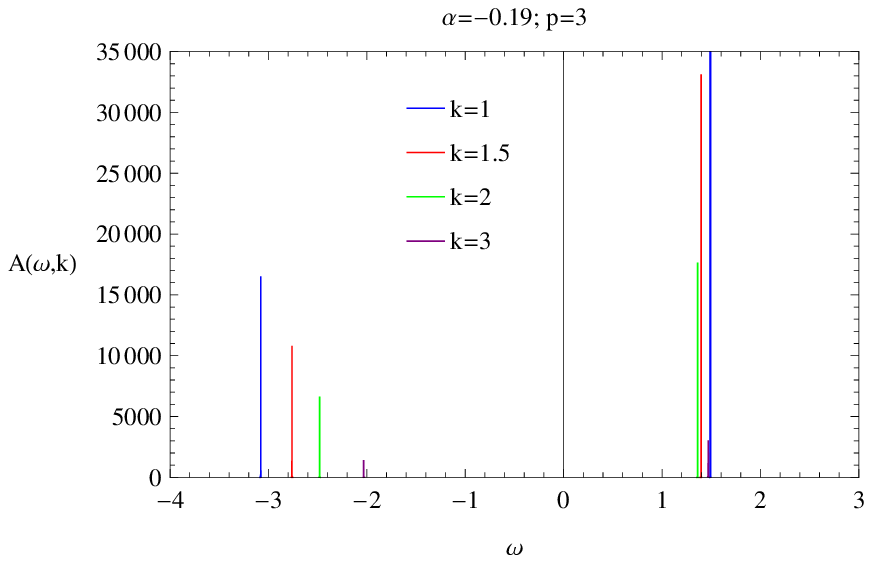}
\includegraphics[scale=0.7]{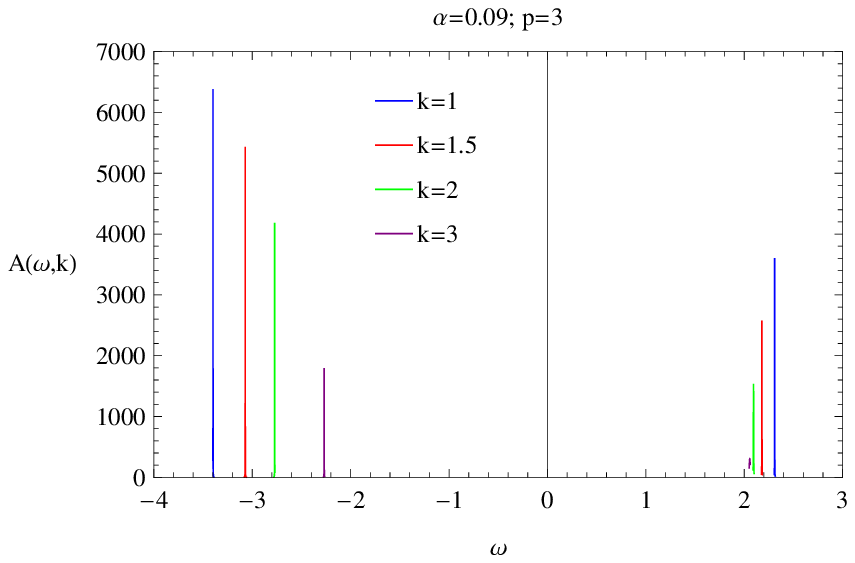}
\caption{\label{figAwnonzeroalpha} $A(\omega,k)$ as function of $\omega$ with sample values of $k$ for $\alpha=-0.19$ and $\alpha=0.09$.}}
\end{figure}

\begin{figure}
\center{
\includegraphics[scale=0.7]{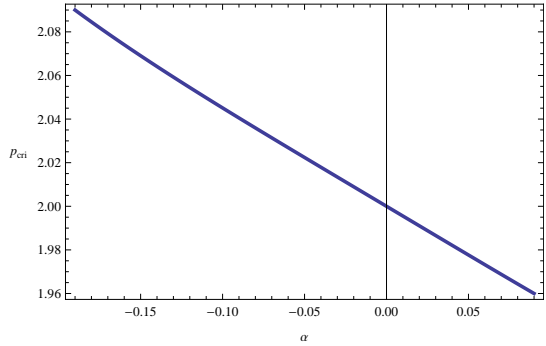}\\
\caption{\label{a-pcri} $p_{cri}$ as a function of $\alpha$.}}
\end{figure}

Now let's turn to discuss the influence of the higher curvature correction on the Fermi gap in the holographic fermion system\footnote{In this subsection, we focus on 5 dimensional gravity background.}. For the nonzero Gauss-Bonnet factor, for example, $\alpha=-0.19$ and $0.09$, we show the 3D plots for $p=3$ in Fig. \ref{fig3D2} where we obtain the gap. The gap in the spectral density exists for all k as shown in Fig. \ref{figAwnonzeroalpha}. In addition, we observe that the critical $p$ for the onset of gap decreases when the Gauss-Bonnet factor becomes bigger by computing the density of state. The explicit relation between $p_{cri}$ and $\alpha$ is shown in FIG. \ref{a-pcri}. It is clear that larger $\alpha$ can promote the effect of $p$. We list the typical values, e.g. $p_{cri}=1.96$ when $\alpha=0.09$, $p_{cri}=2$ when $\alpha=0$ and $p_{cri}=2.09$ when $\alpha=-0.19$. These can also be seen in the inset of Fig. \ref{figgap}. Furthermore, Fig. \ref{figgap} explicitly shows that for the fixed dipole coupling strength, the gap becomes wider with the increase of the Gauss-Bonnet factor.

\begin{figure}
\center{
\includegraphics[scale=0.6]{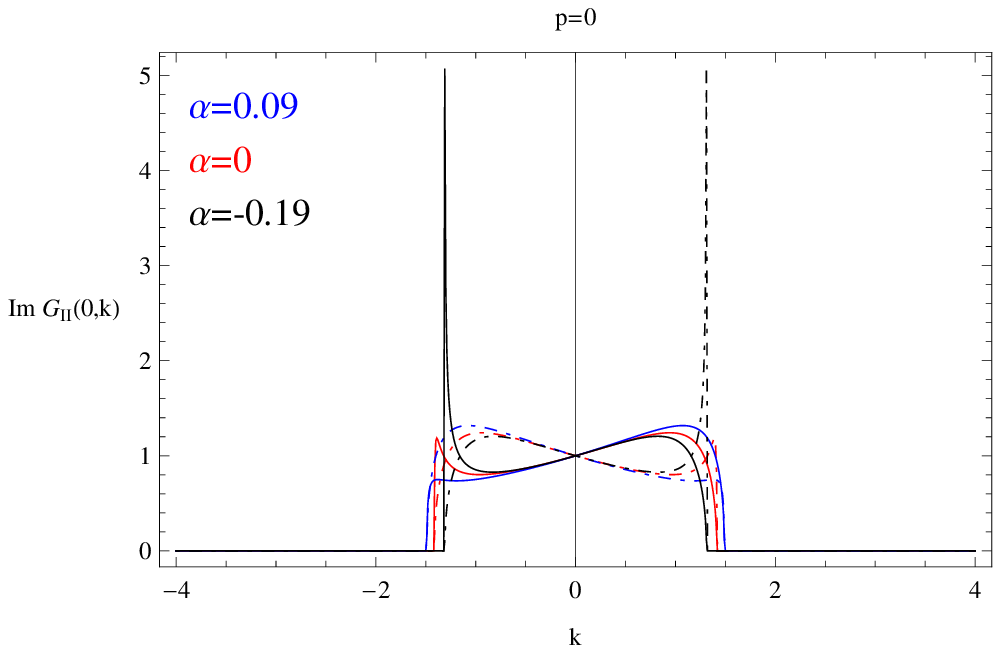}\hspace{1cm}
\includegraphics[scale=0.6]{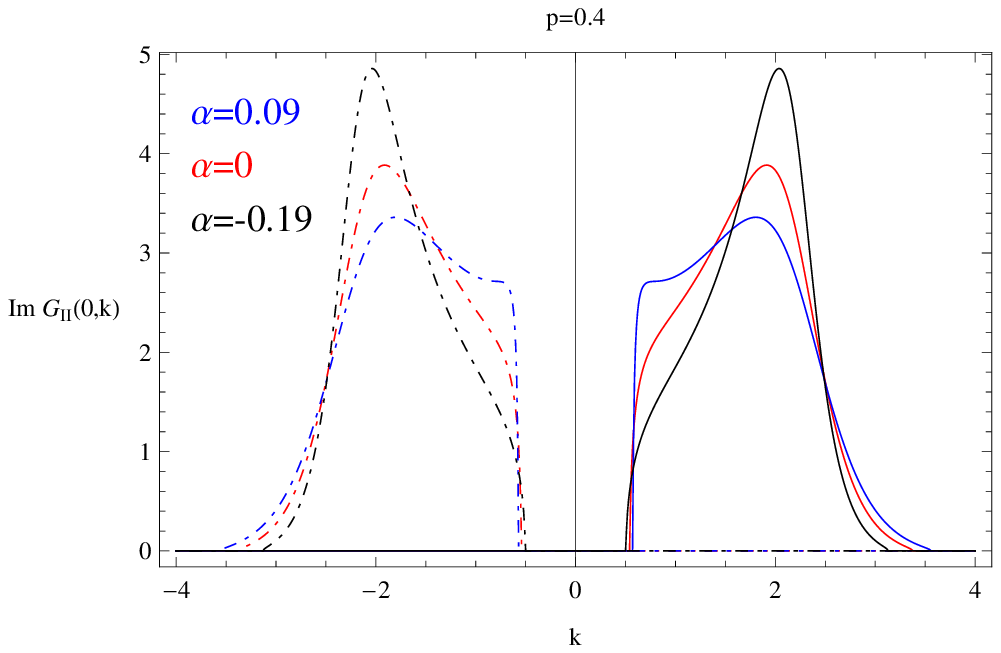}\\
\caption{\label{figkG} $\rm{Im}[G_{11}(0,k)]$(dashed) and $\rm{Im} [G_{22}(0,k)]$(solid) for $p=0$ and $p=0.4$.}}
\end{figure}

Hereafter we report our numerical result in the limit when $\omega=0$. We will pay more attention on the influence in the holographic fermion system by the Gauss-Bonnet factor.  The results are shown in Fig. \ref{figkG}. In the figure, the symmetry of $\rm{Im}[G_{11}(0,-k)]=\rm{Im}[G_{22}(0,k)]$ is clear both for $p=0$ and $p=0.4$. Our numerical results show that both $\rm{Im}[G_{11}(0,k)]$ and $\rm{Im}[G_{22}(0,k)]$ keep nonzero in a range of $k$. This range of $k$ for nonzero $\rm{Im}[G_{11}(0,k)]$ and $\rm{Im}[G_{22}(0,k)]$ at fixed $p$ becomes bigger when the curvature correction $\alpha$ becomes stronger. In the momentum regime for nonzero $\rm{Im}[G_{II}(0,k)]$, $\rm{Im}[G_{II}(\omega,k)]$  become log-oscillatory when $\omega\rightarrow0$\cite{HongLiuNon-Fermi}.  The left plot of Fig. \ref{figkG} shows the log-oscillatory regimes coincide at $p=0$ for all chosen $\alpha$. While this degeneracy shrinks  when we increase the strength of the dipole coupling and breaks down for big enough $p$ in the right plot of Fig. \ref{figkG}.

\begin{figure}
\center{
\includegraphics[scale=0.2]{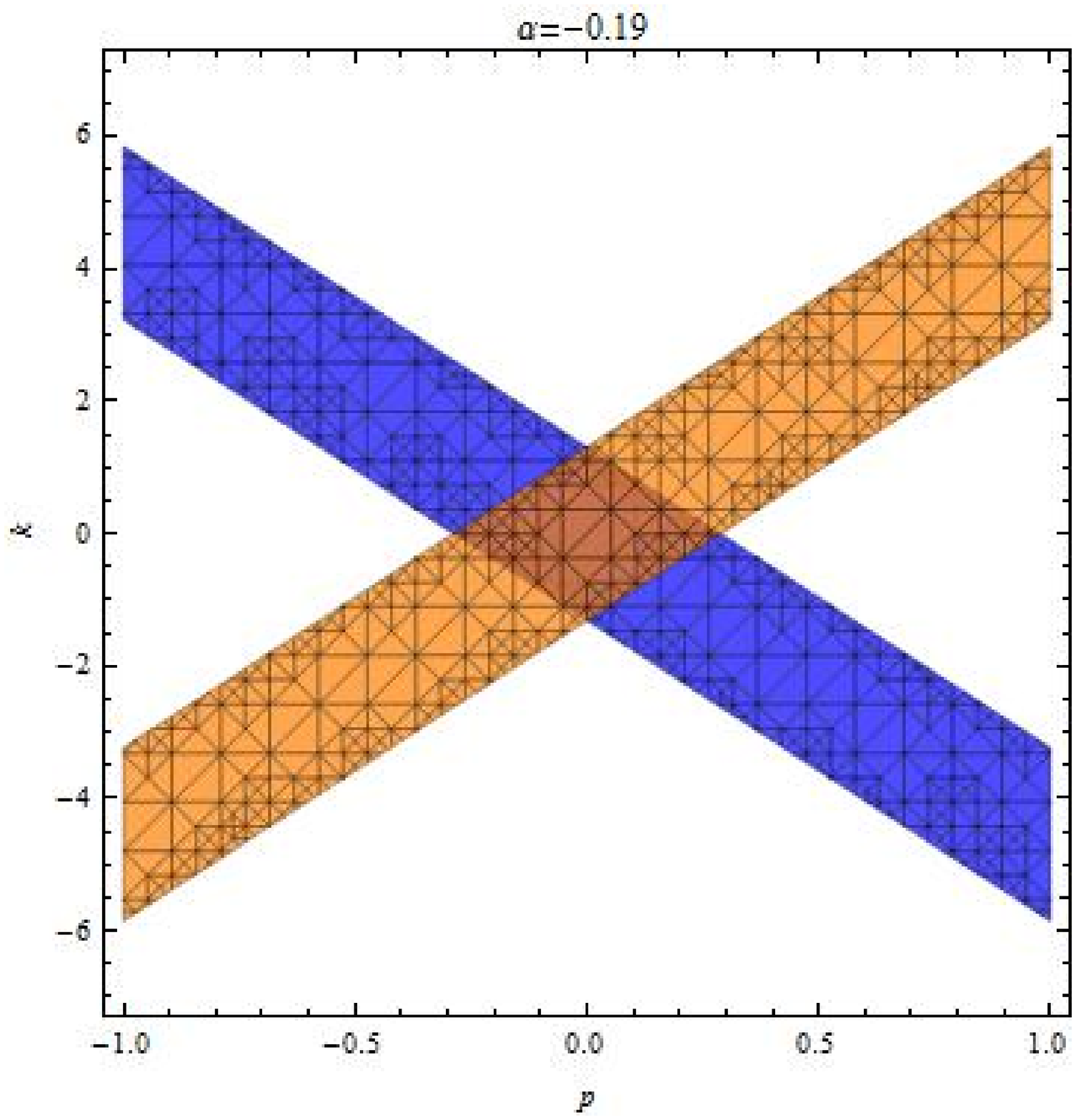}\hspace{1cm}
\includegraphics[scale=0.2]{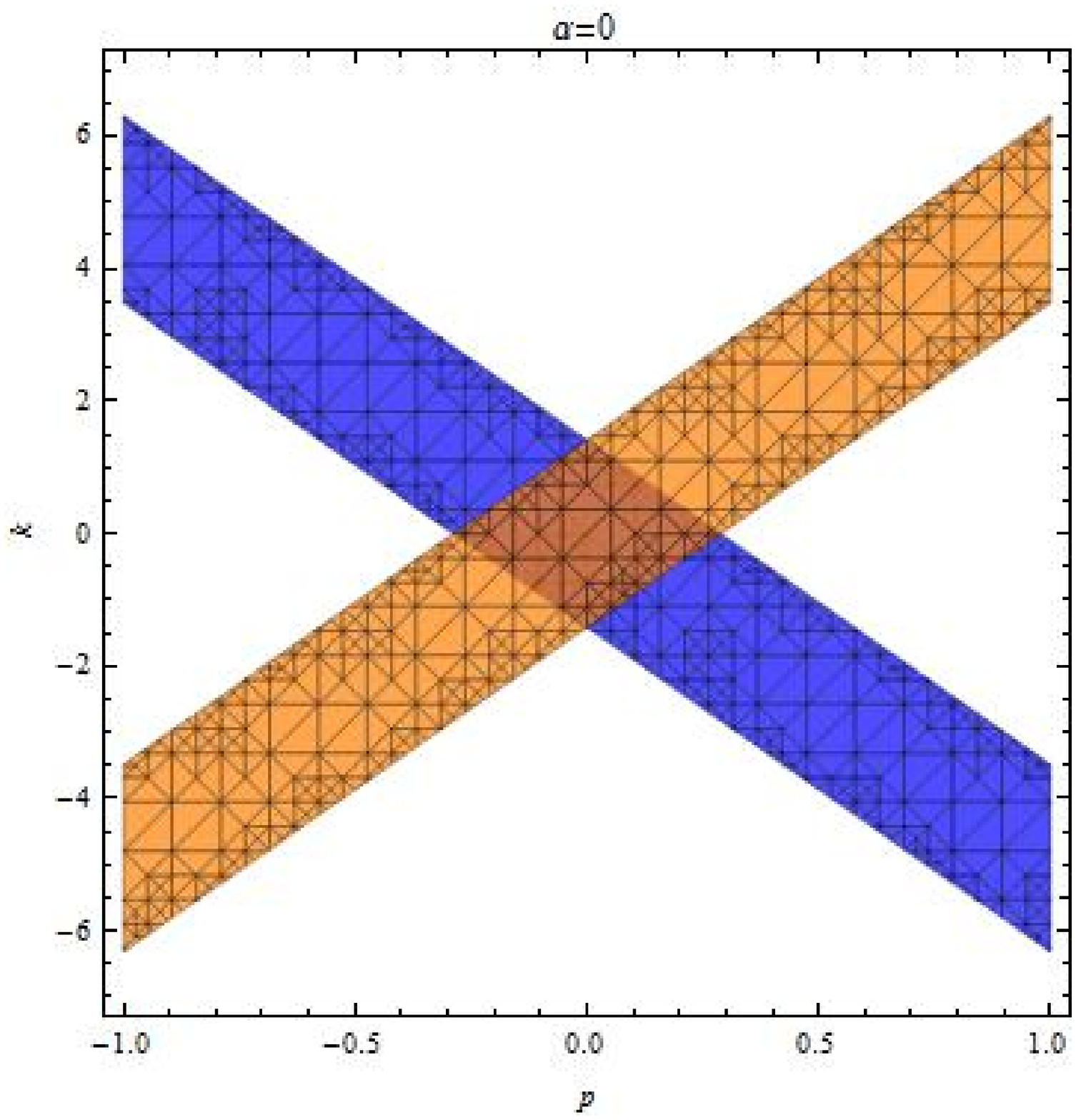}\hspace{1cm}
\includegraphics[scale=0.2]{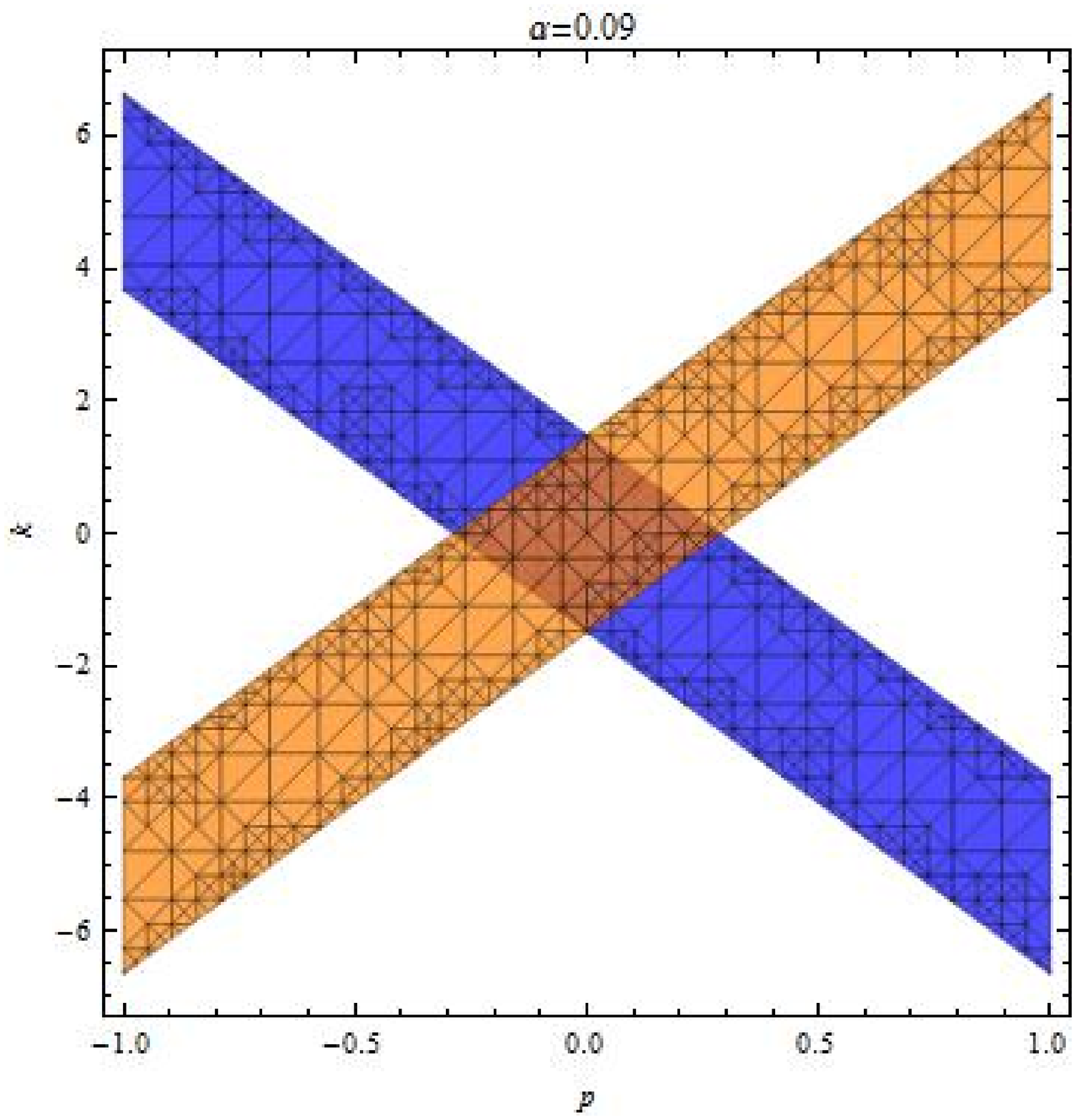}\\
\caption{\label{figpk} Regions of oscillation for various $\alpha$. The orange region is where $\rm{Im}[G_{22}(\omega,k)]$ oscillates while the blue region is where $\rm{Im}[G_{11}(\omega,k)]$ oscillates.}}
\end{figure}

To understand the above numerical result on the log-oscillatory regimes more clearly, we can analyze the $\nu_{I}(k)$ analytically. There is a range of momenta $k\in \mathfrak{I}_{I}$ in which the conformal dimension of the dual $CFT$ operator $\mathbb{O}_{I}$ is imaginary. For $p=0$, the momenta ranges for $\mathbb{O}_{1}$ and $\mathbb{O}_{2}$ are coincident\cite{HongLiuNon-Fermi,HongLiuAdS2}. For $p\neq0$, the momenta ranges for the two operators will separate and the degeneracy will break when $p$ becomes large enough \cite{R.G.Leigh2}. In our model,
the high curvature correction $\alpha$ modifies the conformal dimension in (\ref{nuk}). When $\nu_{I}(k)$ is imaginary for $d=5$, we have
\begin{equation}\label{kosc}
k\in \mathfrak{I}_{I}=\left[\frac{(-1)^{I}2p\mu-\frac{q\mu}{\sqrt{3}}}{L_{\rm eff}},\frac{(-1)^{I}2p\mu+\frac{q\mu}{\sqrt{3}}}{L_{\rm eff}}\right].
\end{equation}
When $k\in \mathfrak{I}_{I}$, we have the imaginary  boundary condition (\ref{bdyw0}), so that $\rm{Im}[G_{II}(0,k)]$ is nonzero when $k\in \mathfrak{I}_{I}$. Considering that $\rm{L_{eff}}$ decreases monotonously as $\alpha$ increases, $\mathfrak{I}_{I}$ will become wider as we increase $\alpha$ when we fix the dipole coupling $p$. This supports the numerical finding above. Furthermore, when $p=0$,  from (\ref{kosc}) it is easy to  obtain $\mathfrak{I}_{1}=\mathfrak{I}_{2}=\mathfrak{I}$ and both $\rm{Im}[G_{11}(\omega,k)]$ for a fixed $\alpha$ and $\rm{Im}[G_{22}(\omega,k)]$ are log-oscillatory. When we turn on $p$, $\mathfrak{I}_{1}$ and $\mathfrak{I}_{2}$ will be separated, so both $\rm{Im}[G_{II}(\omega,k)]$ are oscillatory only when $k\in\mathfrak{I}_{1}\cap\mathfrak{I}_{2}$. When $p$ is increased to $p_c=\frac{1}{2\sqrt{3}}$ which is independent of the value of $\alpha$, $\mathfrak{I}_{1}\cap\mathfrak{I}_{2}=\{0\}$.  Increasing $p$ higher than $p_c$, we have $\mathfrak{I}_{1}\cap\mathfrak{I}_{2}= \emptyset$, which can support the numerical behavior for $p= 0.4$ in Fig. \ref{figkG}. The  separations of the regimes $\mathfrak{I}_{1}$ and $\mathfrak{I}_{2}$ versus $p$ for virous $\alpha$ are presented in Fig. \ref{figpk}.

Now we turn to discuss the case that $\nu_{I}(k)$ is real, in which the boundary conditions (\ref{bdyw0}) at $\omega=0$ are real. Considering that the flow equations (\ref{DiracEF1}) are also real, we conclude that $\rm{Im}[G_{II}(0,k)]=0$
and $\rm{Re}[G_{II}(0,k)]=\frac{a_{I}^{(0)}}{b_{I}^{(0)}}$ from (\ref{coefficients}) and (\ref{greenads2}). So the Fermi momentum $k_{F}$ can be defined as the poles of $G_{II}(0,k)$ with $b_{I}^{(0)}=0$ while $a_{I}^{(0)}$ do not vanish. Taking (\ref{BoundaryBehaviour}) and (\ref{coefficients}) into account and recalling $F_{I} = ( \mathcal{A}_{I} ,  \mathcal{B}_{I})^{T}$, we can deduce directly  $\mathcal{B}_{2}=b_{2}^{(0)} r^{mL}+\cdots$ at $\omega=0$ in the  boundary $r\rightarrow\infty$. To determine $k_{F}$, we need to find the solution to $\mathcal{B}_{2}$ with normalization near the boundary at $\omega=0$.
Setting $\omega=0$, $m=0$ and decoupling the two equation in (\ref{DiracEAB2}), we obtain the equations of $\mathcal{B}_{2}$
\begin{eqnarray} \label{A2B2}
\frac{\sqrt{f(r)}}{v_{-}|_{\omega=0}+k \frac{L_{\rm eff}}{r}}\partial_{r}(\frac{\sqrt{f(r)}\partial_{r}\mathcal{B}_{2}}{-v_{+}|_{\omega=0}+k \frac{L_{\rm eff}}{r}})=\mathcal{B}_{2}.
\end{eqnarray}
Near the horizon, the regular behavior of the field is $\mathcal{B}_{2}(r\rightarrow 1)\sim f(r)^{\frac{\nu_{2}(k)}{2}}$. Near the boundary, we need $\mathcal{B}_{2}(r\rightarrow \infty)=0$ to find the fermi momentum $k_F$.
\begin{figure}
\center{
\includegraphics[scale=0.45]{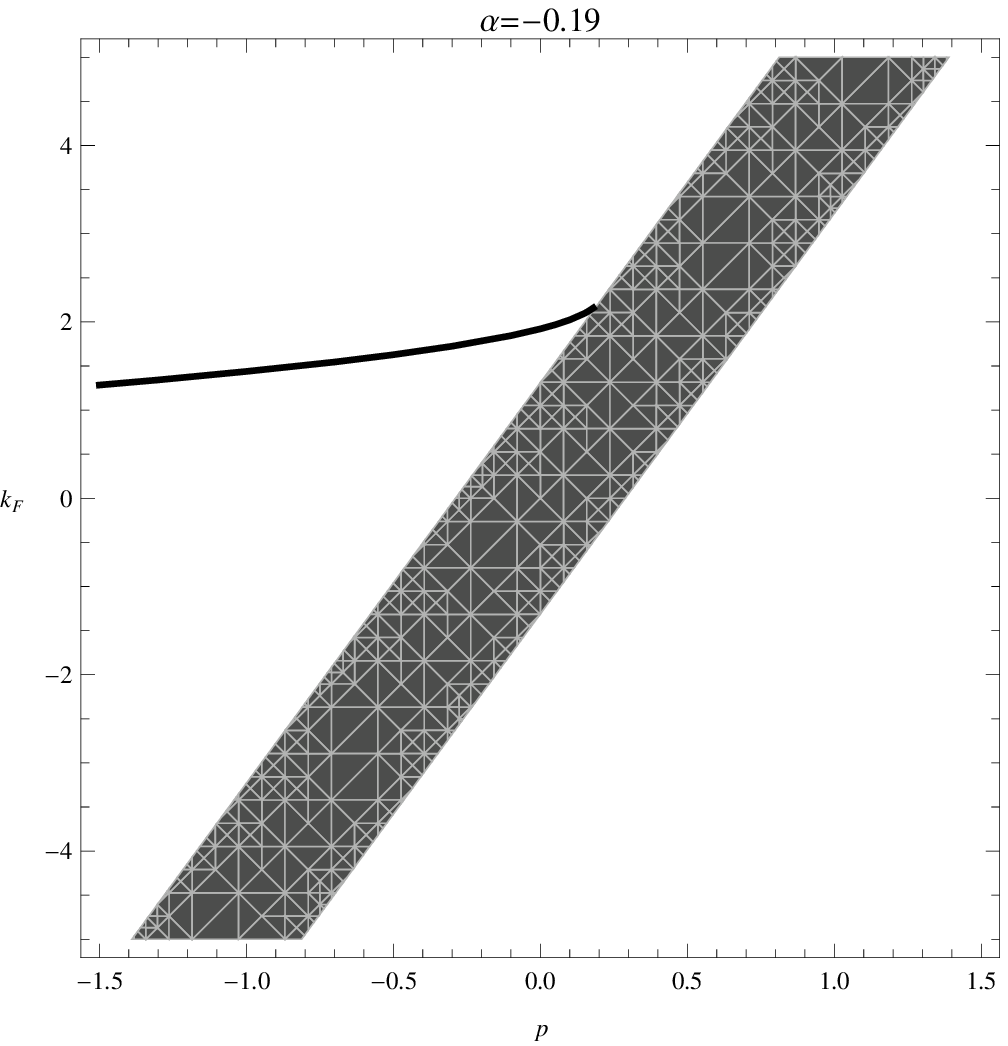}
\includegraphics[scale=0.45]{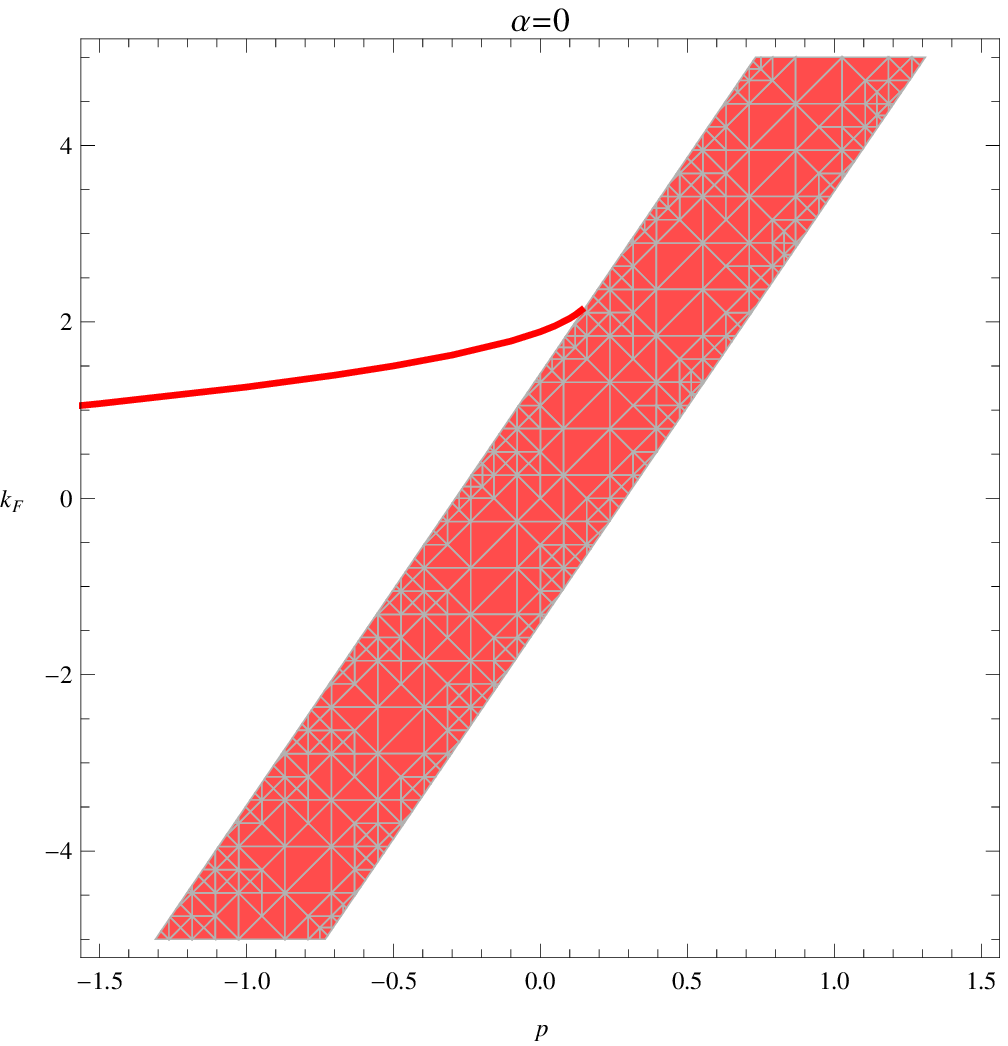}
\includegraphics[scale=0.45]{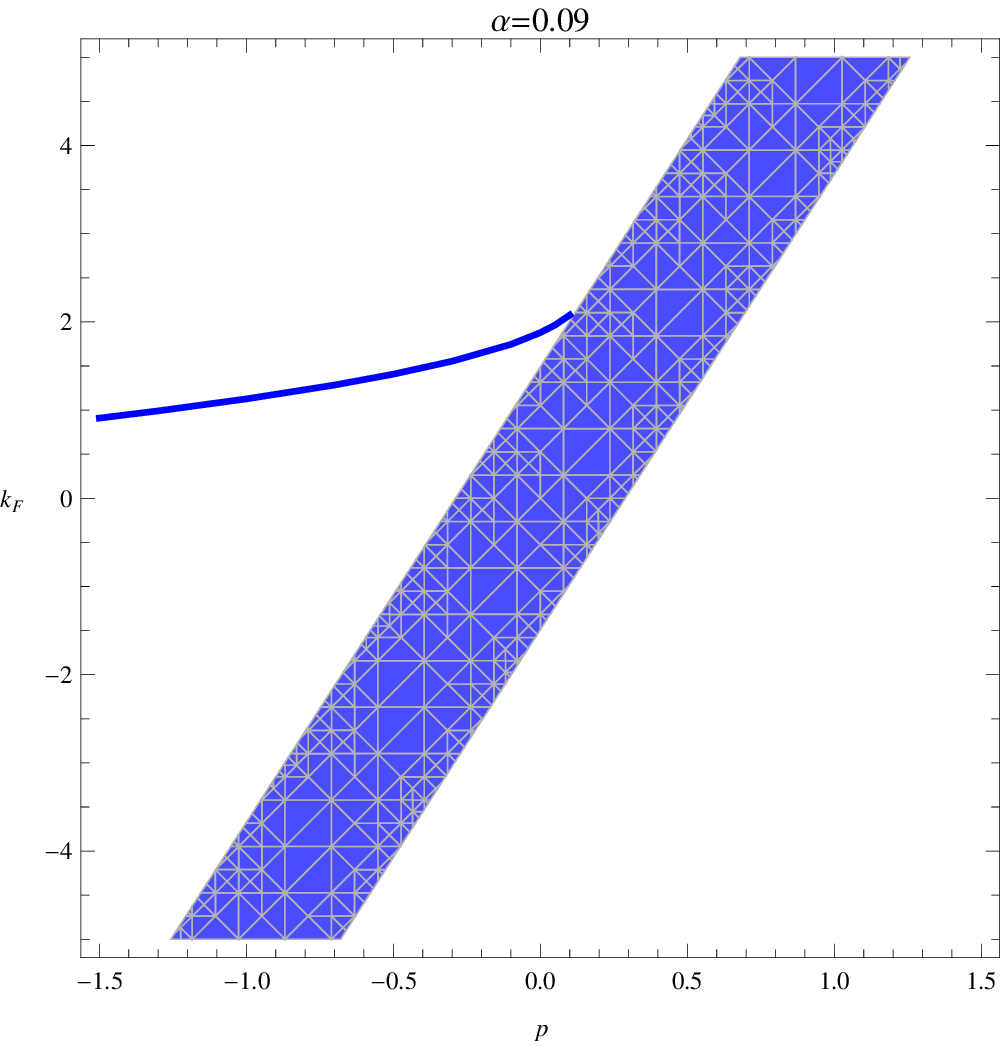}\\
\caption{\label{figkp1} $k_F$ versus p and the log-oscillatory regime $\mathfrak{I}_{2}$.}}
\end{figure}
\begin{figure}
\center{
\includegraphics[scale=0.65]{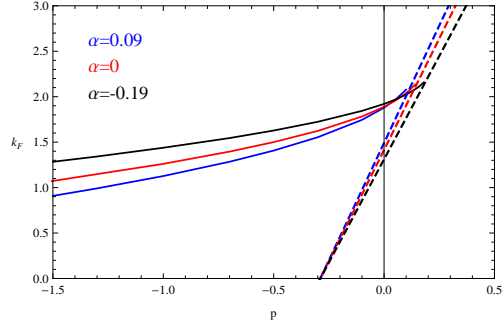}\\
\caption{\label{figkp2} $k_F$ versus p.}}
\end{figure}
By solving the equation (\ref{A2B2}), in Fig. \ref{figkp1}, we show the values of $k_F$ as a function of $p$. The lines show the values of $k_F$ versus $p$ and the orange bands describe the log-oscillatory regions $\mathfrak{I}_{2}$ for the chosen $\alpha$. From each subplot, we see for  the fixed $\alpha$, $k_F$ increases and approaches the boundary of the log-oscillatory regime as we increase $p$. To disclose the influence of the high curvature correction on the Fermi momentum $k_F$, we collect the lines of Fig. \ref{figkp1} into Fig. \ref{figkp2}. We see that for negative $p$, $k_F$ decreases with the increase of $\alpha$. When $p=0$, this dependence of $k_F$ on $\alpha$ was found in  \cite{JPWu1}. But this property changes when $p$ approaches the big enough positive value which still allows the Fermi surface. This may attribute to the steeper boundary of the log-oscillatory regime when $\alpha$ gets bigger as shown by the dashed lines in Fig. \ref{figkp2}.

Besides, in order to explicitly see how $\alpha$ promote the effect of $p$, it is interesting to further investigate the nature of fermion system. We  determine the dispersion relationship via (\ref{LdispersionA}) where the exponent $\nu_2(k=k_F)$ can be calculated through (\ref{nuk}). The typical results are summarized in Table \ref{table2}. We find that when $p$ is negative enough, the excitation near the Fermi surface is always Fermi liquid. If we increase $p$, the excitation will turn to non-Fermi liquid and large enough $p$ will make the Fermi surface disappear. From the table, we can see that lager $\alpha$ will make this turning appear at smaller $p$. This is consistent with the previous result that large $\alpha$ corresponds small $p_{cri}$ as shown in Fig. \ref{a-pcri}.
\begin{center}\label{table2}
\begin{table}[ht]
\begin{tabular}{|c|c|c|c|c|c|c|c|}\hline
& $p=-1$ & $p=-0.1$ & $p=0$ & $p=0.1$ & $p=0.14$ & $p=0.18$ & $p>0.18$ \\ \hline
&$k_F=1.43819$ &$k_F=1.84381$& $k_F=1.92064$& $k_F=2.02279$& $k_F=2.06934$& $k_F=2.15759$& ~\\
$\alpha=-0.19$&$z=1$ &$z=1$& $z=1.14424$& $z=1.86524$& $z=2.77217$& $z=5.80808$& NFS~ \\
& FL & FL & NFL & NFL &  NFL & NFL & ~ \\ \hline
&$k_F=1.26098$ &$k_F=1.78209$& $k_F=1.88730$& $k_F=2.03944$& $k_F=2.13300$& ~& ~\\
$\alpha=0$&$z=1$ &$z=1$& $z=1.38591$& $z=2.73495$& $z=5.64398$& NFS& NFS \\
& FL & FL & NFL & NFL &  NFL & ~ & ~ \\ \hline
&$k_F=0.99100$ &$k_F=1.74514$& $k_F=1.87852$& $k_F=2.07713$& ~& ~& ~\\
$\alpha=0.09$ &$z=1$& $z=1.07353$& $z=1.59722$& $z=3.94973$& NFS& NFS& NFS \\
& FL & NFL & NFL & NFL & ~& ~& ~\\ \hline
\end{tabular}
\caption{\label{table2} the Fermi momentum $k_F$ and the critical exponent $z$ of dispersion relationship for negative, zero and small positive
$p$ for various $\alpha$. NFS means the system doesn't present Fermi surface. FL and NFL denote the excitation near the Fermi surface is Fermi liquid type with $z=1$ and non-Fermi liquid type with $z\neq1$, respectively.}
\end{table}
\end{center}
\section{conclusions and discussions}
We have studied extensively the influences on the holographic fermi system by spacetime dimension and the Gauss-Bonnet factor when there is dipole interaction between fermion and gauge field in the bulk. For the boundary theory dual to the bulk background, we showed that for the higher dimension of the spacetime, the gap starts to emerge in the fermion density of states for the weaker dipole interaction and  the Fermi momentum becomes smaller. Including the higher curvature correction by Gauss-Bonnet factor, we observed that bigger Gauss-Bonnet factor can make the Fermi gap easier to be formed and the gap distance to be enlarged for the fixed nonzero dipole interaction between the fermion and gauge field. Furthermore the bigger Gauss-Bonnet factor can accommodate wider momentum range for the existence of the log-oscillatory, in which Fermi peaks do not occur. The Fermi momentum also changes  when there is Gauss-Bonnet correction in the bulk spacetime. When $p$ is negative, $k_F$ decreases with the increase of the Gauss-Bonnet factor, but this property becomes opposite when $p$ becomes positive enough.

It is important to appreciate the vagaries of holographic studies by reflecting the bulk spacetime influence on the boundary. The next step is natural to ask how the phenomena disclosed due to the introduction of a higher curvature correction and dimensional analysis would complement the physics in superconducting condensate. Another important question is  how much physics the backreaction and the finite temperature will bring to the study. In the future study we will try to answer these questions.

\begin{acknowledgments}
We would like to thank Li-Qing Fang, Xian-Hui Ge, Wei-Jia Li and Hongbao Zhang for their helpful discussion on the related project. In addition, we would like to  thanks Prof. Leigh for pointing out the key point in doing the numerical computation.  X.M Kuang and B. Wang are supported partially by the NNSF of China and the Shanghai Science and
Technology Commission under the grant
11DZ2260700. J.P. Wu
is partly supported by NSFC(No.10975017) and the Fundamental Research Funds for the
central Universities.
\end{acknowledgments}

\end{document}